\documentclass[
twocolumn,
prb,preprintnumbers,amsmath,amssymb]{revtex4}

\usepackage{times}
\usepackage{physics}
\usepackage{graphicx}
\usepackage{amssymb}
\usepackage{amsthm}
\usepackage{amsmath}
\usepackage{dsfont}
\usepackage{bm}
\usepackage{mathrsfs}
\usepackage{bbold}
\usepackage{color}

\def\Xint#1{\mathchoice
   {\XXint\displaystyle\textstyle{#1}}%
   {\XXint\textstyle\scriptstyle{#1}}%
   {\XXint\scriptstyle\scriptscriptstyle{#1}}%
   {\XXint\scriptscriptstyle\scriptscriptstyle{#1}}%
   \!\int}
\def\XXint#1#2#3{{\setbox0=\hbox{$#1{#2#3}{\int}$}
     \vcenter{\hbox{$#2#3$}}\kern-.5\wd0}}

\def\dashint{\Xint-}

\begin{document}

\title{Quantum noise in time-dependent media and cosmic expansion}
\author{Ziv Landau$^{1,2}$ and Ulf Leonhardt$^1$}
\affiliation{
\normalsize{
$^1$Department of Physics of Complex Systems,
Weizmann Institute of Science, Rehovot 761001, Israel}
}

\affiliation{
\normalsize{
$^2$St John's College, University of Cambridge, St John's St., Cambridge CB2 1TP, UK}
}
\date{\today}

\begin{abstract}
In spatially uniform, but time--dependent dielectric media with equal electric and magnetic response, classical electromagnetic waves propagate exactly like in empty, flat space with transformed time, called conformal time, and so do quantum fluctuations. In empty, flat space the renormalized vacuum energy is exactly zero, but not in time--dependent media, as we show in this paper. This is because renormalization is local and causal, and so cannot compensate fully for the transformation to conformal time. The expanding universe appears as such a medium to the electromagnetic field. We show that the vacuum energy during cosmic expansion effectively reduces the weights of radiation and matter by characteristic factors. This quantum buoyancy naturally resolves the Hubble tension, the discrepancy between the measured and the inferred Hubble constant, and it might resolve other cosmological tensions as well. 
\end{abstract}

\maketitle

\section{Introduction}

\begin{figure}[h]
\begin{center}
\includegraphics[width=20pc]{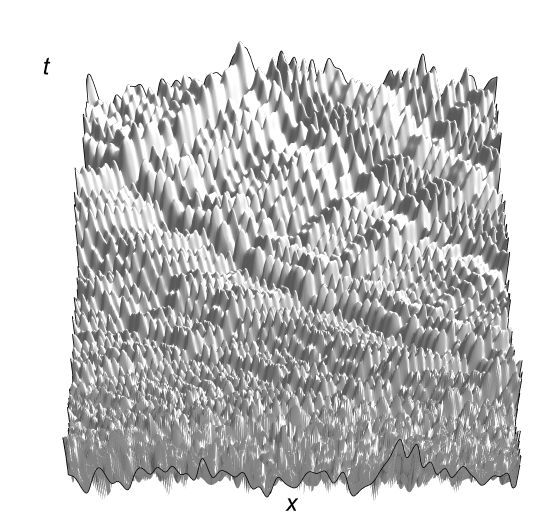}
\caption{
\small{Visualization of quantum noise. Space--time diagram of Gaussian noise in the medium of the expanding universe (for $t$ from $0$ to the present time, with the actual cosmic parameters, Sec.~IVC). Plot of 64 normalized modes summed up with Gaussian random complex amplitudes. One sees how the wavelength changes due to expansion or, equivalently, the time evolution of the medium.
}
\label{fig:noise}}
\end{center}
\end{figure}

Imagine a dielectric medium with time--dependent electric permittivity $\varepsilon$ and magnetic permeability $\mu$. Suppose the medium is made of a spatially uniform, infinitely extended block of material with 
\begin{equation}
\varepsilon = \mu = n(t) \,.
\label{eq:epsmu}
\end{equation}
Note that $n(t)$ describes the refractive index \cite{Jackson}, as $n=\sqrt{\varepsilon\mu}$. Consider, in this medium, the quantum fluctuations of the electromagnetic field \cite{Rodriguez,Buhmann,Forces} (Fig.~\ref{fig:noise}). What is their energy density? What is their gravitational force? Could they influence the expansion of the universe, and if so, how? These are the questions of this paper. As in analogues of gravity \cite{Unruh81,Volovik,Visser,Unsch,Faccionotes,Kolomeisky,Jacquet} they aim at cosmological problems, but are grounded in condensed matter physics, and they do have answers. 

Let us explain step by step. From the wave equation of the electromagnetic vector potential $\bm{A}$ in Coulomb gauge \cite{Jackson,Constants}, $c^2\nabla\times(\mu^{-1}\nabla\times\bm{A})+\partial_t\varepsilon\,\partial_t\bm{A}=0$, follows that $\bm{A}$ satisfies the free--space equation $c^2\nabla\times(\nabla\times\bm{A})+\partial_\tau^2\bm{A}=0$ with the transformed time
\begin{equation}
\tau = \int\frac{dt}{n} \,.
\label{eq:tau0}
\end{equation}
So, like in transformation optics \cite{Leonhardt,Pendry,GREE} the medium of Eq.~(\ref{eq:epsmu}) performs a coordinate transformation of empty space for electromagnetic fields. Similar media have been proposed \cite{McCall} and experimentally demonstrated \cite{Fridman,Lukens} as temporal cloaking devices, or ``history editors'', and time--dependent dielectrics have attracted considerable recent attention \cite{Fan,Segev,Alu}. 

Artificial materials with the properties of Eq.~(\ref{eq:epsmu}) are still difficult to manufacture \cite{SSS}, but one of them occurs in nature: the ``material'' of space (Appendix A). Averaged over cosmological distances ($>$100Mpc) space appears uniform \cite{Survey} and flat \cite{Planck} while expanding with scale factor $a(t)$. Distances measured in wavelengths of electromagnetic radiation grow with the factor $a$ (Fig.~\ref{fig:noise}) as if the wavelength is reduced by the refractive index 
\begin{equation}
n = a(t) \,.
\label{eq:a}
\end{equation}
The electric and the magnetic response of space are the same, and so \cite{Plebanski}, for the electromagnetic field, the scale factor acts like the refractive index of the spatially uniform, time--dependent medium of Eqs.~(\ref{eq:epsmu}) and (\ref{eq:a}). Moreover, this ``dielectric medium'' of space is presumed dispersionless  --- having dielectric properties independent of frequency or wavenumber --- until wavelengths smaller or comparable with the Planck length \cite{Constants}
\begin{equation}
\ell_\mathrm{P} = \sqrt{\frac{\hbar G}{c^3}} = 1.616255 \times 10^{-35}\mathrm{m}\,.
\label{eq:planck}
\end{equation}
This is a consequence of the equivalence principle (Appendix A): space--time acts on everything equally, including every frequency component of the electromagnetic field, until classical general relativity reaches the limit of its validity, which is expected near $\ell_\mathrm{P}$. This fantastic range of frequencies carried without distortion by the medium of space is the reason \cite{Annals} why the quantum noise of the field might be modified, even as the expansion rate, the Hubble parameter \cite{MC}
\begin{equation}
H = \frac{\dot{a}}{a} 
\label{eq:hubble}
\end{equation}
 is as astronomically low as astronomers have measured \cite{Planck,Riess}: $H\approx 70\mathrm{km}/\mathrm{s}/\mathrm{Mpc} \approx 2\times 10^{-18}\mathrm{Hz}$. 
 
What is the quantum state of the electromagnetic field? Since the field propagates like in flat space--time with the transformed time $\tau$ of Eq.~(\ref{eq:tau0}) ---- called conformal time \cite{MC} --- there is a natural ground state of the field called the conformal vacuum \cite{BD}: the ground state of modes made of plane waves with positive frequencies with respect to $\tau$. The universe is filled with the Cosmic Microwave Background (CMB) \cite{MC} and the light of stars {\it etc.}, which is negligible in energy compared with electromagnetic vacuum fluctuations with wavelengths approaching the Planck length. We may thus assume the field to be in the conformal--vacuum state. 
 
 Note that in media where only $\varepsilon$ varies and $\mu$ remains constant, classical electromagnetic waves get reflected \cite{Alu} and amplified \cite{Segev} and quantum particles are produced \cite{Wilson,Hakonen,Veccoli} in a phenomenon \cite{Dodonov,MedoncaBook,Medonca,SchwingerDCE} called the dynamical Casimir effect \cite{SchwingerDCE}. The effect is analogous to the cosmic particle production \cite{BD} of non--conformally invariant fields such as the fluctuations of the gravitational field \cite{Mukhanov} that, presumably, have been amplified from vacuum noise during cosmic inflation \cite{Starobinsky,Guth,Linde}. This amplified noise then got modulated during the formation of the CMB \cite{MC} and provided the seeds for structure formation in the universe \cite{MC}. Analogues of cosmic particle production have been demonstrated \cite{Viermann,Stein} with ultracold atoms. But as we have argued, as long as the electric and the magnetic response of a medium are the same --- Eq.~(\ref{eq:epsmu}) --- the electromagnetic field remains in the conformal vacuum. 
 
 In this paper, we calculate the vacuum energy density of the electromagnetic field in the time--dependent, spatially uniform medium of Eq.~(\ref{eq:epsmu}). We improve on a previous paper \cite{Annals} that relied on the analogy between expanding space and moving media \cite{Annals}. In such media, thermal radiation is produced \cite{GH,FF1,FF2} at the cosmological horizon \cite{Harrison} (where the expansion velocity reaches $c$) similar to the Bekenstein--Hawking radiation of black holes \cite{Bekenstein,Hawking1,Hawking2,Brout} and their analogues \cite{Weinfurtner,Euve,Steinhauer,Drori}. A more careful analysis, conducted here, reveals the limitations of this approach and opens the path to calculating the vacuum energy from first principles.
 
\begin{figure}[h]
\begin{center}
\includegraphics[width=19pc]{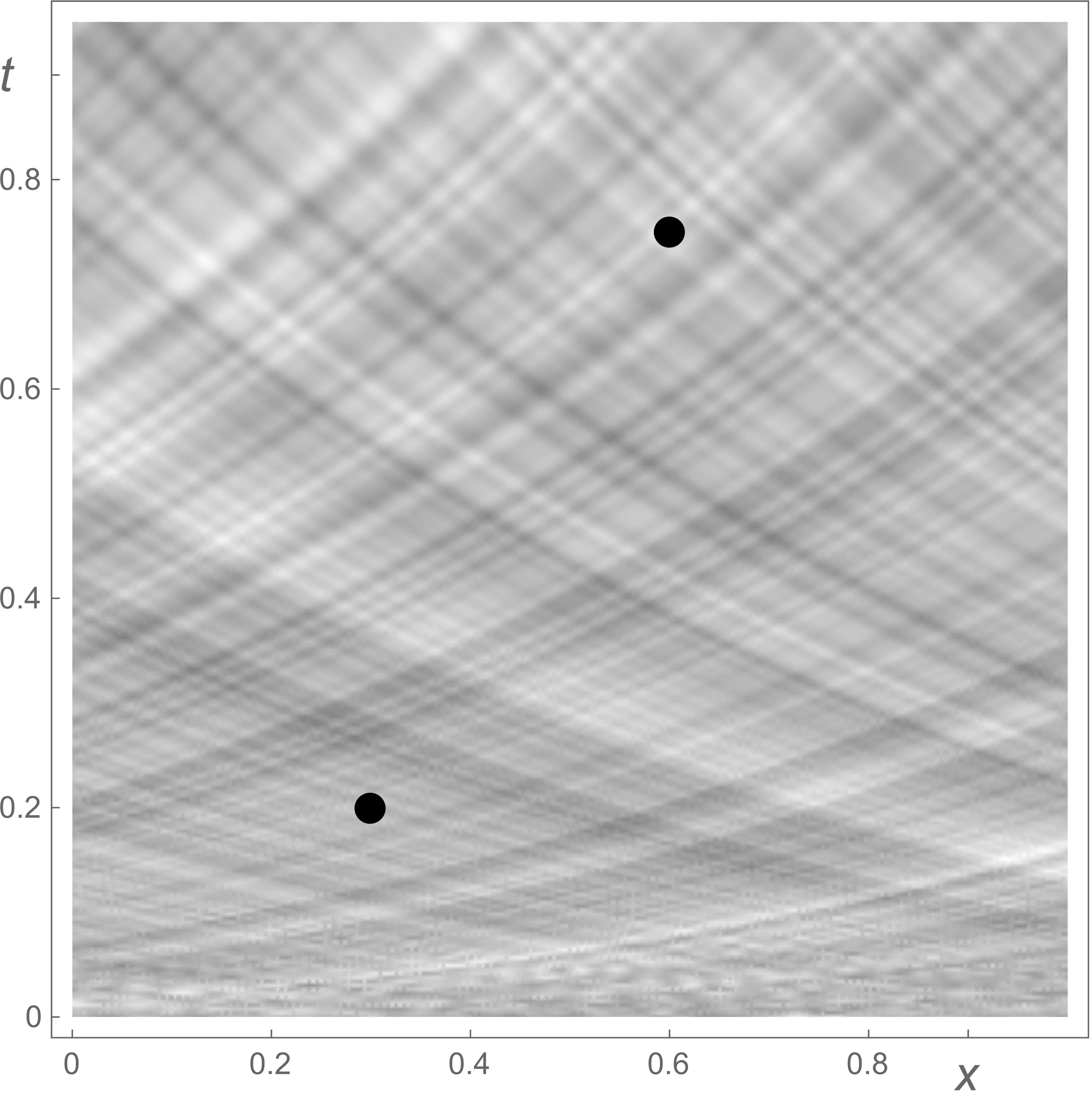}
\caption{
\small{Correlations. Density plot of the same noise as in Fig.~\ref{fig:noise}. Time is measured in units of $1/H_0$ and space in units of $c/H_0$ (Sec.~IVC). Correlations in the noise --- curves of equal amplitudes --- appear along the light cones $s^2=0$ of Eq.~(\ref{eq:s2}): wave noise is organized \cite{LeoBerry}. In Eq.~(\ref{eq:K}) we sample the quantum noise at two space--time points (dots) and calculate the correlation function. 
}
\label{fig:correlation}}
\end{center}
\end{figure}

The classical energy density $u$ of the electromagnetic field is given in terms of the electric field strength $\bm{E}$, the dielectric displacement $\bm{D}$, the magnetic field strength $\bm{H}$ and the magnetic induction $\bm{B}$ as \cite{Jackson} (Appendix A):
\begin{equation}
u = u_\mathrm{E} + u_\mathrm{M} \,,\quad u_\mathrm{E} = \frac{\bm{E}\cdot\bm{D}}{2} \,,\quad u_\mathrm{M} =  \frac{\bm{H}\cdot\bm{B}}{2}
\label{eq:uem}
\end{equation}
with the constitutive equations \cite{Jackson} for the medium of Eqs.~(\ref{eq:epsmu}) and (\ref{eq:a}):
\begin{equation}
\bm{D}=\varepsilon_0 a \bm{E} \,,\quad \bm{B} = \mu_0 a\,\bm{H} \,,\quad \varepsilon_0\mu_0 = c^{-2}
\label{eq:constitutive}
\end{equation}
in SI units where $\varepsilon_0$ denotes the electric permittivity and $\mu_0$ the magnetic permeability of the vacuum. As the fields are derivatives of the vector potential \cite{Jackson}, $\bm{E}=-\partial_t\bm{A}$ (in Coulomb gauge \cite{Jackson}) and $\bm{B}=\nabla\times\bm{A}$, we may regard the energy densities $u_\mathrm{E}$ and $u_\mathrm{M}$ as derivatives of products of two vector potentials at different times and positions $\{t_1,\bm{r}_1\}$ and $\{t_2,\bm{r}_2\}$ and then take the limit
\begin{equation}
t_2\rightarrow t_1 =t\,,\quad \bm{r}_2\rightarrow \bm{r}_1 =\bm{r}\,.
\label{eq:limit}
\end{equation}
We denote the vector potentials at $\{t_1,\bm{r}_1\}$ and $\{t_2,\bm{r}_2\}$  by $\bm{A}_1$ and $\bm{A}_2$, and obtain from Eqs.~(\ref{eq:uem}) and (\ref{eq:constitutive}):
\begin{eqnarray}
u_\mathrm{E} &=& \frac{\varepsilon_0 a}{2}\,\partial_{t_1}\partial_{t_2} \bm{A}_1\cdot \bm{A}_2 \,,
\nonumber\\
u_\mathrm{M} &=& \frac{\varepsilon_0 c^2}{2a} \,(\nabla_1\times\bm{A}_1)\cdot(\nabla_2\times\bm{A}_2) \,.
\end{eqnarray}
In spatially uniform media, the two polarizations of electromagnetic waves are separate and behave identically such that $\bm{A}_1\cdot \bm{A}_2=2A_1 A_2$ and $(\nabla_1\times\bm{A}_1)\cdot(\nabla_2\times\bm{A}_2)=2\nabla_1\cdot\nabla_2 A_1 A_2$ where the $A_1$ and $A_2$ denote one representative  polarization component evaluated at $\{t_1,\bm{r}_1\}$ and $\{t_2,\bm{r}_2\}$. For quantum fields, we need to replace $2A_1A_2$ by the expectation value of the symmetrized field operator products $\widehat{A}_1\widehat{A}_2+\widehat{A}_2\widehat{A}_1$ such that these are Hermitian and their expectation values are real. We thus get for the energies: 
\begin{equation}
u_\mathrm{E} = \frac{\hbar a}{c}\,\partial_{t_1}\partial_{t_2} K \,,\quad
u_\mathrm{M} = \frac{\hbar c}{a} \,\nabla_1\cdot\nabla_2 K
\label{eq:ueum}
\end{equation}
in terms of the correlation function (Fig.~\ref{fig:correlation})
\begin{equation}
K = \frac{\varepsilon_0 c}{2\hbar}\,\langle \widehat{A}_1\widehat{A}_2+\widehat{A}_2\widehat{A}_1\rangle \,.
\label{eq:K}
\end{equation}
Note that such correlation functions have been measured inside materials \cite{Benea,Settembrini}. There, two ultrashort light pulses serve as probes of the quantum noise floor along their world lines $\{t_1,\bm{r}_1\}$ and $\{t_2,\bm{r}_2\}$. In appropriate nonlinear materials, the electromagnetic noise affects the polarizations of the probes. The polarization of each individual probe pulse becomes slightly noisy, but the polarization noise of the two pulses was found to be correlated \cite{Benea}, even outside the light cone \cite{Settembrini}. 

In order to calculate the correlation function $K$ we use a fundamental principle as our starting point, the fluctuation--dissipation theorem \cite{Scheel}. The theorem \cite{Scheel} relates the quantum correlation $K$ to the classical Green function $G$ of the electromagnetic field. This was first done in radar engineering where Rytov \cite{Rytov} calculated the correlations of classical electromagnetic noise from the Green functions. Lifshitz \cite{Lifshitz} and the Landau students Dzyaloshinskii and Pitaevskii extended this connection to quantum correlations and used it to calculate Casimir forces in dielectrics \cite{Lifshitz,DLPrepulsion,DP,DLP}. Here we employ a general form \cite{Annals} of the fluctuation--dissipation theorem in terms of Hilbert transformations \cite{AF} known in physics from the Kramers--Kronig relations \cite{Kronig,Kramers}. One finds (Sec.~II) that the correlation function and energy densities of Eqs.~(\ref{eq:ueum}) and (\ref{eq:K}) tend to infinity when the two times and positions approach each other according to Eq.~(\ref{eq:limit}). But the physical vacuum energies are surely not infinite. In the Casimir effect \cite{Rodriguez}, vacuum fluctuations in spatially varying media exert mechanical forces, and these forces are finite and typically quite small \cite{Atm}. In gravity, if we would take the Planck length of Eq.~(\ref{eq:planck}) as cut--off, the energy density would be in the order \cite{London} of the Planck mass $m_\mathrm{P}$ divided by the cube of the Planck length, with $m_\mathrm{P}=\sqrt{\hbar c/G}\approx 20\mu\mathrm{g}$. Concentrating this macrosopic mass into a Planck--scale volume would change everything we know in physics, and already therefore  must be wrong, despite the empirical fact \cite{Benea,Settembrini} that the bare vacuum correlations of Eq.~(\ref{eq:K}) do exist. How to reconcile these conflicting aspects of vacuum fluctuations? Not all of the vacuum energy can do mechanical work\cite{Rodriguez}. One assumes that the part of the energy density that causes gravitational  forces is the same as the one causing mechanical forces. An experiment is being prepared to test this hypothesis by weighing the quantum fluctuations inside a  material with time--dependent parameters \cite{Kempf,Archimedes}. 

 The procedure of extracting from the vacuum correlations the part that can do mechanical work is called renormalization. Lifshitz renormalization \cite{Lifshitz,DLPrepulsion,DP,DLP} is based on the fluctuation--dissipation theorem \cite{Scheel}: there one subtracts from the Green function $G$ of the full problem the Green function $G_0$ of an infinitely extended, uniform block of material matching the local values of $\varepsilon$ and $\mu$, and then calculates the difference of the corresponding correlation functions $K$ and $K_0$. In our case of spatially uniform, but time--dependent media, Lifshitz renormalization amounts to subtracting from $G$ the Green function $G_0$ in a constant material with $\varepsilon$ and $\mu$ set to $\varepsilon=\varepsilon(t_0)$ and $\mu=\mu(t_0)$ at the time $t_0$ we are interested in. Empirical evidence from Casimir forces in fluids \cite{Munday,Equilibrium} supports the idea that renormalization is local: each local region appears to have its own renormalizer $G_0$. Otherwise the theory would not converge, but the locally renormalized theory \cite{DLPrepulsion} does and it agrees well with the experimental data \cite{Munday,Equilibrium}. 
 
Dzyaloshinskii and Pitaevskii suggested \cite{DP} to use Lifshitz renormalization in spatially inhomogeneous dielectrics. They did not have the computational tools at the time to realise that this is wrong \cite{SimpsonPaper,SimpsonBook}. One might approximate an inhomogeneous medium by a piecewise homogeneous one and make the approximation finer and finer \cite{SimpsonPaper} or build up the medium from the bottom as particles interacting with each other by (retarded) van der Waals forces \cite{Horner}, the continuum limit does not exist, and the force and energy densities do not converge, but approach infinity. It turns out \cite{Itay,Itai} that the renormalizing Green function $G_0$ should not only depend on the local values of the dielectric response functions, but also on their derivatives. Note that the renormalizer must not depend on all local derivatives, because then $G_0$ would be identical with $G$ and nothing were left after renormalization. 

In dispersive media, the minimal number of derivatives for renormalization to converge is two  \cite{Itay,Itai}. In our case, we should make $G_0$ dependent on $a$, $\dot{a}$ and $\ddot{a}$.  This is an absolutely critical assumption, and it was not made in the literature on renormalizing vacuum energies in cosmology \cite{BD}. Another, more obvious assumption was not made either prior to our work \cite{Annals}: causality. The renormalizing Green function can only depend on the local dielectric functions and their derivatives at the earlier of the two times $t_1$ and $t_2$. In our case, if $t_1<t_2$ then $G_0$ should depend on $a(t_1)$, $\dot{a}(t_1)$ and $\ddot{a}(t_1)$ and if  $t_2<t_1$ then $G_0$ should depend on $a(t_2)$, $\dot{a}(t_2)$ and $\ddot{a}(t_2)$. Without dispersion, up to fourth derivatives of the scale factor would be needed to get a converging result after renormalization \cite{Adler}. The resulting energy densities are in the order of $\hbar c$ times the square of the space--time curvature \cite{Wald}. They could only play a role in inflation \cite{Starobinsky,HawkingAnomaly} but not in standard cosmic expansion \cite{MC}. 

Cosmologists have ignored dispersion, but in condensed matter physics it is a natural feature of dielectric media \cite{LL8}. If space appears like a medium for electromagnetic fields, it ought to be dispersive, too \cite{Jacobsondispersion}. Dispersion combined with causality raises the leading order of the vacuum energy density to a level where it matters in cosmic expansion \cite{Annals}. This is because the ``material'' of space  is almost dispersionless until the Planck scale. The renormalization nearly diverges with $\ell^{-2}$ of the cut--off length (Sec.~IIID) and produces \cite{Annals} a vacuum energy of the same order of magnitude as the measured cosmological constant \cite{Planck}. There is still some ambiguity on how the renormalizing Green function depends on $a$, $\dot{a}$ and $\ddot{a}$; we hope to have made the most natural choice (Sec.~IIIB). 

With this renormalizer, we have obtained exceptionally simple results (Sec.~IVB) as follows. The vacuum energy is generated by the cosmic expansion. In turn, the vacuum energy acts back on the expansion by its gravity --- its weight. In cosmology, the matter and energy components in the universe appear as fluids with characteristic equations of state \cite{MC}. For radiation (photons and neutrinos) the energy density $\epsilon_\mathrm{R}$ is proportional to $\hbar\omega/a^3$ with frequency $\omega$ falling with $a$, and so the density $\rho_\mathrm{R}=\epsilon_\mathrm{R}/c^2$ falls with $a^4$. The density $\rho_\mathrm{M}$ of matter (dark and baryonic) falls with $a^3$, while the density $\rho_\Lambda$ of the cosmological constant remains constant \cite{MC}. We will see (Sec.~IV) that the vacuum energy makes gravity partially repulsive, which reduces the weights of the cosmic fluids as if they were becoming buoyant in the vacuum. Remarkably, each fluid turns out to have its own buoyancy. The cosmological constant is unaffected, while the weights $\varrho_\mathrm{R}$ and $\varrho_\mathrm{M}$ of radiation and matter are reduced by characteristic factors:
\begin{equation}
\varrho_\mathrm{R} = \frac{\rho_\mathrm{R}}{1+4\kappa} \,,\quad
\varrho_\mathrm{M} = \frac{\rho_\mathrm{M}}{1+3\kappa} \,,\quad
\varrho_\mathrm{\Lambda} = \rho_\mathrm{\Lambda} \,.
\label{eq:weights}
\end{equation}
Here the parameter $\kappa$ quantifies the influence of the quantum vacuum \cite{Annals}. It is given by the cut--off length $\ell$ versus the Planck length of Eq.~(\ref{eq:planck}) as 
\begin{equation}
\kappa = \frac{8}{9\pi} \, \frac{\ell_\mathrm{P}^2}{\ell^2} \,.
\label{eq:kappa}
\end{equation}
We expect $\ell\gtrsim2\ell_\mathrm{P}$ as $2\ell_\mathrm{P}$ plays the role of a fundamental length scale, for the following reasons. The Planck length of Eq.~(\ref{eq:planck}) is obtained from dimensional analysis \cite{PlanckUnits} and does not have a precise quantitative meaning. However, in Bekenstein's formula \cite{Bekenstein} for the entropy of the black hole, with Hawking's prefactor \cite{Hawking1}, the horizon area divided by $4\ell_\mathrm{P}^2$ gives the entropy. In Jacobson's thermodynamic derivation \cite{Jacobson}  of Einstein's field equations the entropy of causal horizons appears as the area divided by  $4\ell_\mathrm{P}^2$ as well. In both cases, the horizon area is pixelled by $2\ell_\mathrm{P}\times 2\ell_\mathrm{P}$ elements to encode one bit of information each, suggesting that $2\ell_\mathrm{P}$ is indeed a fundamental length. 

For $\ell\gtrsim2\ell_\mathrm{P}$ we obtain another remarkable result, assuming the same value \cite{Planck} for $\rho_\mathrm{M}$ as obtained from measurements \cite{Planck} of the CMB fluctuations \cite{MC}. Strictly speaking, the values of the $\rho_m$ should be modified due to the vacuum corrections of Eq.~(\ref{eq:weights}) in the dynamics of the CMB formation, but we shall argue (Sec.~IVC) that these modifications are likely to be small (perhaps a few percent). What is not small is the correction to the Hubble constant $H_0$. The actual Hubble parameter $H|_{a=1}$ was measured \cite{Riess} to deviate from $H_0$ by a factor of $1.084$, and we match this value for $\ell=4.7\ell_\mathrm{P}$. With our previous model \cite{Annals} we reproduced \cite{Dror} the measurement \cite{Riess} for $\ell=\ell_\mathrm{P}$ but only in perturbation theory. 

The discrepancy between the directly measured and the CMB inferred value of the Hubble constant has been called the Hubble tension \cite{Abdalla} and it has become one of the most actively debated problems of contemporary astrophysics \cite{DiValentino}. While our latest result might still be a coincidence, and while there are in the order of $10^2$ theories explaining the Hubble tension \cite{DiValentino}, our theory is the only one of them that does not require new fields, new modifications of general relativity or new cosmological principles. This does not mean of course that we do not make extrapolations and do not rely on hypotheses  --- but our assumptions are conservative, and we have clearly stated them here. While the Hubble tension and other cosmological tensions \cite{Abdalla} are some of the most urgent problems of astrophysics \cite{DiValentino}, their resolution might very well come from condensed matter physics.

 
 \section{Fluctuation and dissipation}
 
 \subsection{Green functions}
 
 The central quantity to be calculated in this paper  is the correlation $K$ defined in Eq.~(\ref{eq:K}) and evaluated in the conformal vacuum. The energy densities are obtained in Eq.~(\ref{eq:ueum}) by differentiating $K$ with respect to the space--time coordinates $\{t_1,\bm{r}_1\}$ and $\{t_2,\bm{r}_2\}$. In the fluctuation--dissipation theorem \cite{Scheel}  we relate the correlation $K$ of the fluctuations to the dissipation $\Gamma$ defined as 
\begin{equation}
\Gamma = \frac{\varepsilon_0 c}{2i\hbar}\,\langle \widehat{A}_1\widehat{A}_2-\widehat{A}_2\widehat{A}_1 \rangle = \frac{\varepsilon_0 c}{2i\hbar}\, \big[\widehat{A}_1,\widehat{A}_2] \,.
\label{eq:gamma}
\end{equation}
 We removed the quantum average, as the commutator of the field is state--independent \cite{Annals}. The commutator gives the retarded and advanced Green functions $G_\pm$ such that \cite{Annals} 
\begin{equation}
\Gamma = \frac{1}{2c}\,(G_+-G_-) \,.
\label{eq:dissipation}
\end{equation}
The retarded Green function $G_+$ describes a classical light pulse emitted from a point source at $\{t_1,\bm{r}_1\}$ with unit strength, propagating in $\{t_2,\bm{r}_2\}$ coordinates. The advanced Green function $G_-$ describes in $\{t_2,\bm{r}_2\}$ coordinates how a pulse is focused such that it becomes absorbed at the point $\{t_1,\bm{r}_1\}$. The difference $G_--G_+$ satisfies the homogeneous wave equation without source or drain, and so it  is neither absorbed nor emitted, but focused \cite{Fink}. In our case, Eqs.~(\ref{eq:epsmu}) and (\ref{eq:a}), the field propagates as in Minkowski space with respect to conformal time, and so we have \cite{Jackson}
\begin{equation}
G_\pm = -\frac{1}{4\pi r} \,\delta(\tau\mp r/c) 
\label{eq:G}
\end{equation}
in terms of the radius
\begin{equation}
r = |\bm{r}_2-\bm{r}_1| 
\label{eq:r}
\end{equation}
and the conformal--time difference 
\begin{equation}
\tau = \int_{t_1}^{t_2} \frac{dt}{a} \,.
\label{eq:tau}
\end{equation}
In time--independent media, $G_\pm$ would only depend on the time difference $t_2-t_1$. There $G_-$ is the time reverse of $G_+$ and so the Fourier transform $\widetilde{G}_-(\omega)$ equals $\widetilde{G}_+(-\omega)$, which implies $\widetilde{G}_--\widetilde{G}_+=2i\mathrm{Im}\widetilde{G}_+$ that corresponds to the local density of states \cite{DoS}.

\subsection{Fluctuation--dissipation theorem}

How is the dissipation $\Gamma$ related to the correlation $K$? We see from Eqs.~(\ref{eq:K}) and (\ref{eq:gamma}) that $K$ is the real part and $\Gamma$ the imaginary part of a complex function $f$:
\begin{equation}
K=\mathrm{Re} f \,,\quad \Gamma = \mathrm{Im} f \quad\mbox{with}\quad f=\frac{\varepsilon_0 c}{\hbar}\,\langle \widehat{A}_1\widehat{A}_2 \rangle \,.
\label{eq:f}
\end{equation}
The function $f$ is often called the Wightman function \cite{BD}. Let us analyze its properties. The field operator $\widehat{A}$ can be decomposed into a superposition of modes \cite{Essential}:
\begin{equation}
\widehat{A} = \sum_k \left(\widehat{a}_kA_k + \widehat{a}_k^\dagger A_k^*\right)
\label{eq:superposition}
\end{equation}
with the quantum annihilation and creation operators $\widehat{a}_k$ and $\widehat{a}_k^\dagger$ and the classical mode functions $A_k$. In our case, the mode functions can be taken as plane waves oscillating with positive frequencies \cite{Essential} $\omega_k$ with respect to conformal time. In the conformal vacuum, where $\widehat{a}_k|0\rangle = 0$, the Wightman function appears as a superposition of terms oscillating as $\exp(-i\omega_k\tau)$ in the conformal--time difference, Eq.~(\ref{eq:tau}). As these are analytic functions decaying on the lower half plane of $\tau$ the Wightman function $f$ is also analytic and decaying for $\mathrm{Im}\,\tau<0$. From Cauchy's theorem \cite{AF} follows that the real part and the imaginary part on the real axis, Eq.~(\ref{eq:f}),  are connected by the Hilbert transformation \cite{AF}:
\begin{equation}
K(\tau) = -\frac{1}{\pi} \dashint_{-\infty}^{+\infty} \frac{\Gamma(\tau_0)}{\tau_0-\tau}\,d\tau_0  \,.
\label{eq:hilbert}
\end{equation}
This is the fluctuation--dissipation theorem \cite{Scheel}  for our case of time--dependent media with equal electric and magnetic response: the correlation $K$ is given in terms of the dissipation $\Gamma$ by the Hilbert transformation (\ref{eq:hilbert}) with respect to the conformal--time difference (\ref{eq:tau}). Connections between real and imaginary part are familiar in physics from the Kramers--Kronig relations \cite{Kronig,Kramers} between the real and the imaginary part of the Fourier--transformed susceptibility $\widetilde{\chi}(\omega)$. There causality requires that $\chi(t)=0$ for $t<0$, which implies that $\widetilde{\chi}(\omega)$ is analytic and decaying on the upper half $\omega$ plane. This gives the same relation as Eq.~(\ref{eq:hilbert}) for $\mathrm{Re}\widetilde{\chi}$ and $\mathrm{Im}\widetilde{\chi}$, apart from a different sign in the prefactor, because in the underlying Cauchy theorem \cite{AF} the integration contour runs mathematically--positive (counter--clockwise) on the upper half plane and negative (clockwise) on the lower half plane. 

One might object that the imaginary part of $f$, the dissipation $\Gamma$ of Eqs.~(\ref{eq:dissipation}) and (\ref{eq:G}), consists of delta functions that are not mathematical functions. However, they are associated with poles: the functions $f_+= i\int_0^\infty e^{i\omega\tau}\,d\omega$ and $f_-= -i\int_0^\infty e^{-i\omega\tau}\,d\omega$ are analytic on the upper (+) or lower (-) half plane, respectively, and the integrals give $f_\pm=\tau^{-1} \pm i\pi \delta(\tau)$. In the fluctuation--dissipation theorem, delta functions in $\Gamma$ get translated into poles in $K$. If these delta functions were negative they would still produce the same poles, but should be read as analytic on the lower half plane, which will become important later on (Sec.~IIIC). 

To proceed, we insert Eqs~(\ref{eq:dissipation}) and (\ref{eq:G}) in Eq.~(\ref{eq:hilbert}) and obtain:
\begin{equation}
K=-\frac{1}{(2\pi)^2 s^2}
\label{eq:Kresult}
\end{equation}
in terms of the Minkowski metric
\begin{equation}
s^2 = c^2\tau^2-r^2 \,.
\label{eq:s2}
\end{equation}
We see that the vacuum correlations are singular at the light cone ($s=0$). We also see that the correlations persist outside of the light cone (where $s^2<0$) in agreement with experiment \cite{Settembrini}. They change sign: fields are correlated outside and anti--correlated inside the light cone. These correlations arise, because the randomness in wave noise stems from the coefficients $\widehat{a}_k$ and $\widehat{a}_k^\dagger$ in the mode decomposition of Eq.~(\ref{eq:superposition}) whereas the mode functions are deterministic waves extending over space and time. Wave noise is organized \cite{LeoBerry} (Fig.~\ref{fig:correlation}).

\subsection{de Sitter space}

An instructive example is the case of exponential expansion called de Sitter space \cite{deSitter,LeviCivita} where the Hubble parameter of Eq.~(\ref{eq:hubble}) is constant:
\begin{equation}
H = \mathrm{const}
\label{eq:deS}
\end{equation}
and so $a(t)=a_0\,e^{Ht}$ with $a_0=\mathrm{const}$. Let us express the two times $t_1$ and $t_2$ of the Green function and correlation function in terms of their average $t$ and difference $\theta$:
\begin{equation}
t =\frac{1}{2}\,(t_1+t_2) \,,\quad \theta = t_2-t_1 \,.
\label{eq:times}
\end{equation}
We obtain for the difference (\ref{eq:tau}) in conformal time:
\begin{equation}
\tau = \frac{2}{aH}\,\mathrm{sinh}\,x
\label{eq:taudeS}
\end{equation}
with $a=a(t)$ and 
\begin{equation}
x = \frac{H\theta}{2} \,.
\label{eq:xdeS}
\end{equation}
Suppose we replace $\theta$ by the complex
\begin{equation}
\theta_* = \theta - i\beta \quad \mbox{with}\quad \beta = \frac{2\pi}{H} \,.
\end{equation}
We see from Eqs.~(\ref{eq:taudeS}) and (\ref{eq:xdeS}) that $\tau(\theta_*)$ is real but changes sign. We obtain from Eqs.~(\ref{eq:Kresult}) and (\ref{eq:s2}) that $K(\theta_*)=K(\theta)$ and from Eqs.~(\ref{eq:dissipation}) and (\ref{eq:G}) that $\Gamma(\theta_*)=-\Gamma(\theta)$, and so
\begin{equation}
f(\theta_*)=f^*(\theta) \,.
\label{eq:KMS}
\end{equation}
This is the Kubo--Martin--Schwinger relation \cite{Annals,Scheel} that characterizes a thermal state with temperature 
\begin{equation}
k_\mathrm{B}T = \frac{\hbar H}{2\pi}
\label{eq:GH}
\end{equation}
where $k_\mathrm{B}$ denotes Boltzmann's constant. The Kubo--Martin--Schwinger relation (\ref{eq:KMS}) suggests that the vacuum in the exponentially expanding universe appears as thermal radiation. Let us compare the de Sitter correlation function $K_\mathrm{deS}$ with the correlation $K_\mathrm{th}$ of thermal fluctuations in Minkowski space, with inverse temperature $\beta=\hbar/(k_\mathrm{B}T)$. One obtains via conformal mapping \cite{Annals}:
\begin{equation}
K_\mathrm{th} = - \frac{1}{8\pi^2 c^2\rho^2}\,\partial_\rho \ln\left(e^{H\theta}-e^{H\rho}\right)\left(e^{H\theta}-e^{-H\rho}\right)
\label{eq:thermal}
\end{equation}
with $\rho=r/c$. For $r\rightarrow 0$ one gets indeed $K_\mathrm{deS}=K_\mathrm{th}$ with the temperature of Eq.~(\ref{eq:GH}) for $\beta=2\pi/H$. This is the Gibbons--Hawking effect \cite{GH}: an observer at rest with the exponentially expanding universe perceives the quantum vacuum as thermal relation. However, for $r\neq 0$ ({\it i.e.} $\bm{r}_1\neq\bm{r}_2$) the correlations functions of the vacuum in de Sitter space and the thermal correlation function of Eq.~(\ref{eq:thermal}) are different. A nonlocal detector might not see thermal radiation at all or thermal radiation with a different temperature. Here is an intriguing example \cite{VolovikdeS}: take the ionization of an atom as detection event of Gibbons--Hawking radiation. Note that the atom is not a point detector, as the ionized electron must get away from it, ideally to infinity. Use the coordinates $\bm{x}=a\bm{r}$ where the exponential expansion appears as a stationary flow \cite{CC}. In these coordinates, the electron tunnels out of the atom through a potential barrier set by the ionization energy and gradually lowered with $|\bm{x}|$ due to cosmic expansion. The tunnel rate suggests \cite{VolovikdeS} that the atom detects thermal radiation of twice the temperature of Eq.~(\ref{eq:GH}). We interpret this beautiful result \cite{VolovikdeS} as a feature of nonlocal detection: the expansion of the universe has helped the electron tunnel out of the atom to infinity, which raises the Gibbons--Hawking temperature, remarkably by exactly a factor of two \cite{VolovikdeS}. 

\subsection{Energy densities}

Return to the general case of arbitrary time evolution $a(t)$. We wish to compute the electromagnetic energy densities that according to Eq.~(\ref{eq:ueum}) depend on the correlation function $K$ given [Eqs.~(\ref{eq:Kresult}) and (\ref{eq:s2})] by the conformal time $\tau$ defined in Eq.~(\ref{eq:tau}). We express the two times $t_1$ and $t_2$ in terms of the time difference $\theta$ and average $t$ as
\begin{equation}
t_1 = t - \frac{\theta}{2} \,,\quad t_2 = t + \frac{\theta}{2} \,.
\label{eq:times2}
\end{equation}
If we replace $\theta$ by $-\theta$ the times $t_1$ and $t_2$ are exchanged, and the defining integral (\ref{eq:tau}) of $\tau$ changes sign: $\tau$ is an odd function of $\theta$. For small time differences $\theta$, we could regard the scale factor in Eq.~(\ref{eq:tau}) as a constant and get $\tau\sim\theta/a$. 

Let us work out the higher terms in the power series of $\tau(\theta)$. We shall do this in a form that will be useful later on (Sec.~III) for renormalization. We express the scale factor in terms of the Hubble parameter $H(t)$ of Eq.~(\ref{eq:hubble}):
\begin{equation}
a(t) = a(t_1)\exp\left(\int_{t_1}^t H\,dt \right)
\label{eq:aint}
\end{equation}
and expand $H(t)$ in a power series around $t_1$:
\begin{equation}
H = \sum_{m=0}^\infty \frac{H^{(m)}}{m!}\,(t-t_1)^m \,.
\label{eq:Hseries}
\end{equation}
The next non--vanishing order in $\tau(\theta)$ we obtain from the first--order term in the power series (\ref{eq:Hseries}). We work out the integral (\ref{eq:tau}) as a power series in $\theta$ and get
\begin{equation}
\tau \sim \frac{\theta}{a} +\frac{H^2-\dot{H}}{24a} \,\theta^3 \,.
\label{eq:tauex}
\end{equation}
Higher--order terms can be obtained from higher orders in the Hubble series (\ref{eq:Hseries}) but expression (\ref{eq:tauex}) is already sufficient for calculating all divergent orders of the electromagnetic energy densities; higher orders just produce converging terms, as we will see next. 

Formula (\ref{eq:ueum}) for the energy densities contains the derivatives $\partial_{t_1}\partial_{t_2}$ and $\nabla_1\cdot\nabla_2$. The time derivatives we express as $\partial_{t_1} = -\partial_\theta+\frac{1}{2}\partial_t$ and $\partial_{t_2} = \partial_\theta+\frac{1}{2}\partial_t$ according to Eq.~(\ref{eq:times}). As $K$ is only a function of $r=|\bm{r}_2-\bm{r}_1|$ we have for the spatial derivatives $\nabla_1\cdot\nabla_2=-\nabla^2=-\partial_r^2-\frac{2}{r}\partial_r$. We thus get from Eq.~(\ref{eq:ueum}):
\begin{eqnarray}
u_\mathrm{E} &=& -\frac{\hbar a}{c}\,\left(\partial_\theta^2-\frac{1}{4}\,\partial_t^2\right)K \,,\nonumber\\
u_\mathrm{M} &=& -\frac{\hbar c}{a} \left(\partial_r^2+\frac{2}{r}\partial_r\right) K \,.
\label{eq:ueum2}
\end{eqnarray}
These expressions involve maximally second derivatives of $K$ with respect to $\theta$ while $K$ is given by Eqs.~(\ref{eq:Kresult}) and (\ref{eq:s2}). Higher--order terms of $\tau(\theta)$ will thus produce converging results in our limiting procedure [Eq.~(\ref{eq:limit})] that corresponds to the limits $\theta\rightarrow 0$ and $r\rightarrow 0$. In this paper we are solely interested in the divergencies, because only they can produce energies that matter on a cosmological scale for a cut--off on the Planck scale. 

Note that the $u_\mathrm{E}$ and $u_\mathrm{M}$ are coordinate densities, not proper densities. During the cosmic expansion the volume element grows with $a^3$ and so we must divide the coordinate densities by $a^3$ to get the proper densities (Appendix A): 
\begin{equation}
\epsilon_\mathrm{E} = \frac{u_\mathrm{E}}{a^3} \,,\quad \epsilon_\mathrm{M} = \frac{u_\mathrm{M}}{a^3} \,.
\label{eq:energies}
\end{equation}
These energy densities diverge in the limits $\theta\rightarrow 0$ and $r\rightarrow 0$. Diverging expressions may depend on the order of limits. If we take the limit $r\rightarrow 0$ first we get 
\begin{eqnarray}
\epsilon_\mathrm{E} &\sim& \frac{3\hbar}{2\pi^2 c^3\, \theta^4} - \frac{\hbar\,(2H^2+\dot{H})}{8\pi^2c^3\,\theta^2}\,, \nonumber\\
\epsilon_\mathrm{M} &\sim& \frac{3\hbar}{2\pi^2 c^3\, \theta^4} - \frac{\hbar\,(H^2-\dot{H})}{4\pi^2c^3\,\theta^2}\,,
\end{eqnarray}
whereas if we take the limit $\theta\rightarrow 0$ first we have 
\begin{equation}
\epsilon_\mathrm{E}  = \epsilon_\mathrm{M} = -\frac{\hbar c}{2\pi^2\ell^4}
\label{eq:barenergies}
\end{equation}
in terms of the proper length 
\begin{equation}
\ell = a r \,.
\label{eq:length}
\end{equation}
In the latter case (taking the limit $\theta\rightarrow 0$ first) the energy densities do not depend on the Hubble parameter $H$ at all. Note that this is an exact result valid for arbitrary $\ell$, not only in the limit $\ell\rightarrow 0$. So for $\theta\rightarrow 0$ we get exactly the same energy densities as in empty Minkowski space: no trace is left of Gibbons--Hawking radiation. This illustrates again that the Gibbons--Hawking effect \cite{GH,LeoBerry} requires $r\rightarrow 0$: a local observer at rest with the universe.

In our case, the different orders of limits encode different types of dispersion, as follows. We describe the dispersion by a cut--off, a smallest time or a smallest length. For ordinary dielectrics, the order of magnitude of  the smallest time corresponds to the lifetime of the highest resonance, whereas the smallest length corresponds to the molecular or atomic size. Typical resonances have decay rates in the order of MHz and atoms are as small as a few $10^{-10}\mathrm{m}$. They are therefore much longer in time than in space, and so for ordinary dielectrics we should take the limit $r\rightarrow 0$ first. For the ``dielectric'' of space the nature of dispersion is not known yet. Following Jacobson \cite{Jacobsondispersion} we assume {\it spatial dispersion} where the medium of space--time consists of elements that are shorter in time than in space such that we take the limit $\theta\rightarrow 0$ first. We shall see (Sec.~III) that in this case we will get particularly simple and transparent expressions for the renormalized energies. 

 \section{Renormalization} 
 
 \subsection{Preamble}
 
 The previous version \cite{Annals} of the theory of quantum noise in the expanding universe has relied on the analogy between the expansion and moving media \cite{CC} that suggests the thermal radiation of cosmological horizons \cite{EPL}. Horizons can radiate with excellent approximations to thermal Planck spectra \cite{Ziv}. However, we have seen (Sec.~IIC) that even for the simplest case --- exponential expansion --- the quantum vacuum appears as thermal radiation only to local observers at rest, but not in the full correlation function of the electromagnetic field. While we are going to apply many of the ideas and the same mathematical technique as in the previous theory \cite{Annals} we need a different physical starting point. Let us begin at the beginning. 
 
 \subsection{Hubble truncation}
 
Renormalization is local. For our case of time--dependent media, the renormalizing Green function $G_0$ should only depend on the scale factor $a$ and its first and second derivatives $\dot{a}$ and $\ddot{a}$ at the time of emission \cite{Itay,Itai}, which according to Eq.~(\ref{eq:hubble}) is equivalent to the scale factor $a$ and the Hubble parameter $H$ with its first derivative $\dot{H}$. For $G_0$ we thus take $a$ from Eq.~(\ref{eq:aint}) and truncate the Hubble series (\ref{eq:Hseries}) at the first order:
\begin{equation}
H = H(t_1)+\dot{H}(t_1)(t-t_1) \quad\mbox{for}\quad t_2>t_1
\label{eq:H1}
\end{equation}
when $t_1$ is the time of emission and $t_2$ the time of reception. Consider the conformal--time difference $\tau$ that determines the Green functions in Eq.~(\ref{eq:G}). Following the same procedure as before (Sec.~IID) we calculate $\tau$ as a power series in the actual time difference $\theta$:
\begin{equation}
\tau \sim \frac{\theta}{a} +\frac{H^2-\dot{H}}{24a} \,\theta^3 + \frac{\ddot{H}}{24a} \,\theta^4 \quad\mbox{for}\quad \theta>0 \,.
\label{eq:taurenorm1}
\end{equation}
Due to the truncation of the Hubble series we have got a fourth--order term, but $\tau$ must be an odd function of $\theta$. If we reverse the order of times and take $t_2$ as the time of emission and $t_1$ as the time of reception such that 
\begin{equation}
H = H(t_2)+\dot{H}(t_2)(t-t_2) \quad\mbox{for}\quad t_1>t_2
\label{eq:H2}
\end{equation}
we replace $\theta$ by $-\theta$ [Eq.~(\ref{eq:times2})] and get 
\begin{equation}
\tau \sim \frac{\theta}{a} +\frac{H^2-\dot{H}}{24a} \,\theta^3 - \frac{\ddot{H}}{24a} \,\theta^4  \quad\mbox{for}\quad \theta<0  \,.
\label{eq:taurenorm2}
\end{equation}
Indeed, $\tau(\theta)$ is an odd function, but it is no longer analytic \cite{AF}: it has a discontinuity in the fourth derivative. Consequently, for the correlation function $K$ we must not use expressions (\ref{eq:Kresult}) and (\ref{eq:s2}), because the fluctuation--dissipation theorem (Sec.~IIB) relies on analyticity. The correlation function $K$ is the real part and the dissipation $\Gamma$ the imaginary part of the Wightman function, Eq.~(\ref{eq:f}). The dissipation $\Gamma$ is proportional to the difference between the retarded and the advanced Green functions. For analytic functions, the real and imaginary parts are connected, but not for arbitrary functions with discontinuities in derivatives. For finding the correlation $K_0$ of the renormalizer we thus need to re--express the problem in terms of analytic functions. 

\begin{figure}[h]
\begin{center}
\includegraphics[width=20pc]{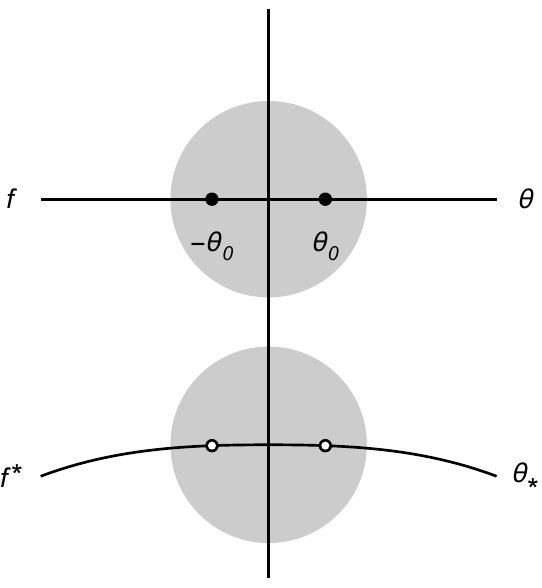}
\caption{
\small{Kubo--Martin--Schwinger relation. We consider the complex time--difference $\theta$ for $\theta\sim 0$ and $\theta\sim-2\pi i/H$ (light disks) and assume condition (\ref{eq:KMS}) for the Wightman function $f$. The dots indicate the delta--function singularities of the dissipation (\ref{eq:dissipation}) with Green function (\ref{eq:G}) at $\pm\theta_0$. On the $\theta_*$ curve they change sign (black to white).
}
\label{fig:KMS}}
\end{center}
\end{figure}

Here we will make another assumption on our renormailzation procedure that follows the same spirit as the truncation of the Hubble series, Eqs.~(\ref{eq:H1}) and (\ref{eq:H2}). There we are fitting the evolution $a(t)$ of the scale factor with nearly--exponential pieces at each moment of time: with local de Sitter spaces. Here we assume that each local de Sitter space also obeys a Kubo--Martin--Schwinger relation \cite{Annals}. Specifically, we express the conformal--time difference $\tau$ of Eqs.~(\ref{eq:taurenorm1}) and (\ref{eq:taurenorm2}) in the form of Eq.~(\ref{eq:taudeS}) with $a=a(t)$ and $H=H(t)$. Instead of the $x$ of Eq.~(\ref{eq:xdeS}) we obtain from the series (\ref{eq:taurenorm1}) and (\ref{eq:taurenorm2}): 
\begin{equation}
x = \frac{H\theta}{2\eta}
\label{eq:x}
\end{equation}
with 
\begin{equation}
\eta \sim 1 + \frac{\dot{H}}{24}\theta^2 \mp \frac{\ddot{H}}{24}\theta^3 \,.
\label{eq:eta}
\end{equation}
So far, this is simply an equivalent re--expression of the series (\ref{eq:taurenorm1}) and (\ref{eq:taurenorm2}) around $\theta\sim 0$. Now, if we replace the $\theta$ in Eq.~(\ref{eq:x}) with 
\begin{equation}
\theta_* = \theta - \frac{2\pi i\eta}{H}
\label{eq:thetastar}
\end{equation}
such that 
\begin{equation}
x = \frac{H\theta_*}{2\eta}
\label{eq:xstar}
\end{equation}
we get $\tau(\theta_*)=-\tau(\theta)$ and therefore $\Gamma(\theta_*)=-\Gamma(\theta)$ from  Eqs.~(\ref{eq:dissipation}) and (\ref{eq:G}). We assume that $K(\theta_*)=K(\theta)$ such that for the Wightman function (\ref{eq:f}) the Kubo--Martin--Schwinger relation (\ref{eq:KMS}) holds. This is the assumption we make (Fig.~\ref{fig:KMS}).

 \subsection{Geometrization}
 
Assumption (\ref{eq:KMS}) has a great mathematical advantage: it turns an analytic problem into a geometrical one (Fig.~\ref{fig:KMS}). Instead of a real function with discontinuous fourth derivative, we have a curve in the complex plane to discuss, the curve $z=\theta_*$ parametrized by $\theta$ according to Eqs.~(\ref{eq:eta}) and (\ref{eq:thetastar}). If we manage to re--parameterize this curve with an analytic function $z(w)$ for real $w$ we have solved our problem: we have found an analytic representation for the Wightman function along the curve $\theta_*$ and, via the Kubo--Martin--Schwinger relation (\ref{eq:KMS}), on the real axis as well. This is where we need analyticity for calculating the correlation $K_0$ from the Green functions via the fluctuation--dissipation theorem. 

\begin{figure}[h]
\begin{center}
\includegraphics[width=18pc]{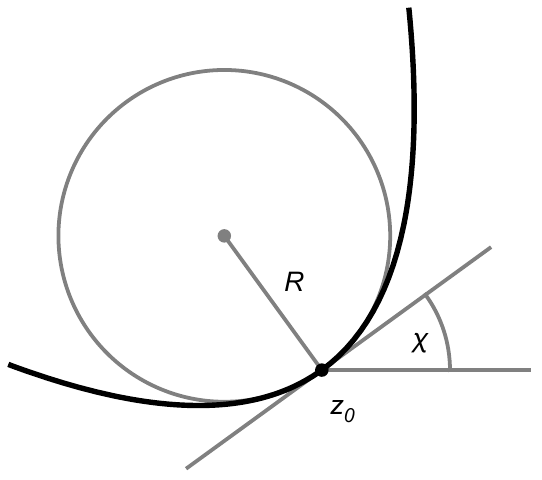}
\caption{
\small{Curve in the complex plane, going through the point $z_0$. Near $z_0$ we characterize the curve by its parameter--invariant geometric properties: the angle $\chi$ [Eq.~(\ref{eq:angle})] and the radius of curvature $R$ [Eq.~(\ref{eq:R})]. The next, more abstract, invariant quantity is the curvature change [Eqs.~(\ref{eq:d2}) and (\ref{eq:schwarz})] (not shown here). 
}
\label{fig:curve}}
\end{center}
\end{figure}

Consider a curve in the complex $z$ plane going through a point $z_0$ (Fig.~\ref{fig:curve}). We only need to discuss this curve in the vicinity of $z_0$. In order to translate one parametrization into another we use the invariants of the curve (Fig.~\ref{fig:curve}) as these are parameter--independent geometrical quantities \cite{Annals}. The first invariant is the infinitesimal length, the line element 
\begin{equation}
dl=|dz|=|z'|\,dw \,.
\label{eq:dl}
\end{equation}
Here and elsewhere the prime denotes the derivative of $z$ with respect to the parameter, here $w$. The second invariant is the angle $\chi$ of the curve:
\begin{equation}
\chi = \mathrm{arg} z' = \mathrm{Im}\ln z' \,.
\label{eq:angle}
\end{equation}
The third invariant is the radius of curvature $R$ with \cite{Annals}
\begin{equation}
\frac{1}{R} = \frac{d\chi}{dl} = \frac{1}{|z'|}\,\mathrm{Im}\, \frac{z''}{z'} \,.
\label{eq:R}
\end{equation}
We are also going to use the fourth invariant, the curvature change \cite{Annals}:
\begin{equation}
\frac{d^2\chi}{dl^2} = \frac{1}{|z'|^2}\,\mathrm{Im}\, \{z,w\} 
\label{eq:d2}
\end{equation}
expressed \cite{Annals} in terms of the Schwarzian derivative \cite{AF}
\begin{equation}
\{z,w\} \equiv \left(\frac{z''}{z'}\right)' -\frac{1}{2}\left(\frac{z''}{z'}\right)^2.
\label{eq:schwarz}
\end{equation}
For the curve $z=\theta_*$ given by Eqs.~(\ref{eq:eta}) and (\ref{eq:thetastar}) with parameter $\theta$ we have
\begin{equation}
z_0= -\frac{2\pi i}{H} \,,\quad
\chi_0 = 0 \,,\quad
\left.\frac{d\chi}{dl} \right|_0 = - \frac{\pi}{6} \,\frac{\dot{H}}{H} 
\label{eq:invariants}
\end{equation}
and the discontinuous
\begin{equation}
\left.\frac{d^2\chi}{dl^2} \right|_{\pm0} = \pm \frac{\pi}{2} \,\frac{\ddot{H}}{H} \,.
\label{eq:invariant4}
\end{equation}
We wish to parameterize the same curve with an analytic function $z(w)$ having the same invariants. This function must have a branch point that matches the discontinuity in Eq.~(\ref{eq:invariant4}) for the fourth invariant, Eq.~(\ref{eq:d2}). The branch points of analytic functions are power--law or logarithmic \cite{AF}. Here we need the logarithm.

The imaginary part of the logarithm smoothly goes from zero to $+i\pi$ on the upper half plane; it jumps from zero to $+i\pi$ on the real axis if we take the logarithm as analytic on the upper half plane. On the lower half plane it goes or jumps from zero to $-i\pi$. On which half plane are we? Near $\theta_*$ where we parameterize our curve, the poles of the Wightman function $f$ have reversed sign (Fig.~\ref{fig:KMS}). In order to match them with poles in Cauchy's theorem, we need to go around the poles clock--wise (mathematically negative) which means we are on the lower half plane (see also Sec.~IIB). We require \cite{Annals}
\begin{equation}
\{z,w\} = \gamma + \delta \ln w 
\label{eq:log}
\end{equation}
with real constant $\delta$ and complex constant $\gamma$ to be determined from the invariants, Eqs.~(\ref{eq:invariants}) and (\ref{eq:invariant4}). For simplicity, we also require $z'(0)=1$. We obtain from Eqs.~(\ref{eq:d2}), (\ref{eq:schwarz}) and (\ref{eq:log}) that $d^2\chi/dl^2|_{+0} = \mathrm{Im}\gamma$ and $d^2\chi/dl^2|_{-0} = \mathrm{Im}\gamma - \pi\delta$ (on the lower half plane). As $d^2\chi/dl^2|_{-0}=-d^2\chi/dl^2|_{+0}$ we have $\pi\delta = 2\,d^2\chi/dl^2|_{+0}$ and hence from Eq.~(\ref{eq:invariant4}):
\begin{equation}
\delta = \frac{\ddot{H}}{H} \,.
\label{eq:delta}
\end{equation}
We solve Eq.~(\ref{eq:log}) for $w\sim 0$ and obtain:
\begin{equation}
z \sim z_0 + w + i\beta w^2 + \frac{\delta}{6}\,w^3\ln w + Q\,w^3
\label{eq:solution}
\end{equation}
with $Q=-\beta^2+\frac{\gamma}{6} - \frac{11}{36}\,\delta$ and the real constant $\beta$ (not to be confused with the inverse temperature). We might fit the constants $\beta$ and $\gamma$ to our invariants (\ref{eq:invariants}) but, as we will see next, the only important quantity is $\delta$. With this we have solved our mathematical problem: we have found the analytic  parameterization (\ref{eq:solution}) of the curve $z=\theta_*$. 

\subsection{Renormalized energy density}

In renormalization, causality causes discontinuities in the conformal--time difference, expressed in Eq.~(\ref{eq:eta}). These discontinuities prevent the direct application of the fluctuation--dissipation theorem for calculating the correlation function from the Green functions. We have extended the actual time difference $\theta$ to the complex $z$ plane and considered the curve where the Kubo--Martin--Schwinger relation (\ref{eq:KMS}) holds. There we have transformed $z$ to $w$ such that the discontinuity in the third derivative of the curve is described by an analytic function with a logarithmic branch point, Eq.~(\ref{eq:solution}). Now we return to the real axis. We use the Kubo--Martin--Schwinger relation (\ref{eq:KMS}) and the analyticity of $z(w)$ to calculate the correlation of the renormalizer. We do this only to leading order, as all other terms will have negligible energies. 

On the real axis $\theta=\mathrm{Re}\, z$. There we obtain from Eq.~(\ref{eq:solution}) $\theta \sim w + \frac{\delta}{6}\,w^3\ln w$ neglecting the cubic term. Inverting this relationship: 
\begin{equation}
w \sim \theta - \frac{\delta}{6}\,\theta^3\ln |\theta| \quad\mbox{for}\quad \theta\sim 0 
\label{eq:wtheta}
\end{equation}
and real $\theta$ gives the transformed time $w$ for which we then can apply the fluctuation--dissipation theorem. We transform the Green function from $\tau$ to $w$ and get from Eq.~(\ref{eq:G}):
\begin{equation}
G_{\pm0} = -\frac{1}{4\pi r} \left|\frac{dw}{d\tau}\right|_{w_\pm} \delta(w-w_\pm)
\label{eq:G0} 
\end{equation}
where the $w_\pm$ denote the zeros of $\tau=\pm r/c$ as a function of $w$. In leading order, $\tau\sim\theta/a$ and so $w_\pm=w(\pm ra/c)$ with the function $w(\theta)$ given by Eq.~(\ref{eq:wtheta}). As this is an odd function we have
\begin{equation}
w_\pm\sim\pm w(\theta_0) \,,\quad\theta_0=\frac{ar}{c} \,.
\label{eq:wpm} 
\end{equation}
We also have $dw/d\tau\sim a w'(\theta)$ and $w'(-\theta)=w'(\theta)$. The Green function $G_{\pm 0}$ of Eq.~(\ref{eq:G0}) generates the dissipation $\Gamma_0$ according to Eq.~(\ref{eq:dissipation}). Seen as a function of $w$, the dissipation is the imaginary part of an analytic function. The real part --- the correlation $K_0$ --- we obtain by replacing the delta functions in $\Gamma_0$ by poles:
\begin{equation}
K_0 = -\frac{a\,|w'(\theta_0)|}{8\pi^2 cr} \left(\frac{1}{w-w_+} - \frac{1}{w-w_-}\right) .
\label{eq:K0}
\end{equation}
Now turn to calculating the energies. We insert formula (\ref{eq:K0}) with definitions (\ref{eq:wtheta}) and (\ref{eq:wpm}) in Eqs.~(\ref{eq:ueum2}) and (\ref{eq:energies}) and take the limit $\theta\rightarrow 0$. This gives the energy densities of the renormalizer as a function of the cut--off $r$ that we express as the proper length $\ell = ar$. We obtain up to order $\ell^{-2}$:
\begin{equation}
\epsilon_{\mathrm{E}0}  \sim \epsilon_{\mathrm{M}0} \sim -\frac{\hbar c}{2\pi^2\ell^4} +\frac{\hbar c}{2\pi^2\ell^2} \,\frac{\delta}{6c^2} \,.
\label{eq:renenergies}
\end{equation}
Higher--order terms make negligible contributions to the energy. The dominant term  of order $\ell^{-4}$ matches exactly the bare energy densities, Eq.~(\ref{eq:barenergies}). In renormalization, by taking the difference between $\epsilon$ and $\epsilon_0$ it removes the contribution that disagrees \cite{Weinberg} with the real vacuum energy \cite{Planck} by 120 orders of magnitude for $\ell$ in the order of the Planck length $\ell_\mathrm{P}$. The term of order $\ell^{-2}$ comes from causality in renormalization. We obtain from Eq.~(\ref{eq:delta}) for the sum of the electric and magnetic energy densities of Eq.~(\ref{eq:renenergies}):
\begin{equation}
\epsilon_\mathrm{vac}=\epsilon-\epsilon_0  \sim -\frac{\hbar c}{\pi^2\ell^2} \,\frac{\ddot{H}}{6c^2 H} \,.
\label{eq:total}
\end{equation}
This energy density is of exactly the right order of magnitude to influence the expansion of the universe (Sec.~IV). But it is not the only vacuum energy present, as we shall see next. 

\subsection{Anomaly} 

The first law of thermodynamics requires that $dE=-pdV$ for an adiabatic process, such as the cosmic expansion, that does not change the entropy, $dS=0$. Here $V$ denotes the volume of an arbitrary region of space, $p$ is the pressure and $E$ the energy in this volume. The energy is given by the energy density $\epsilon$ times the volume, which implies $d\epsilon=-(\epsilon + p)\,dV/V$. During cosmic expansion $V$ grows with $a^3$ and so we have $dV/V=3\,da/a$. We thus obtain \cite{LL2} from the first law and the definition (\ref{eq:hubble}) of the Hubble parameter:
\begin{equation}
\dot{\epsilon} = - 3(\epsilon + p) H \,.
\label{eq:f2}
\end{equation}
This is one of Friedmann's equations \cite{Friedmann1,Friedmann2}. Note that this simple consequence of thermodynamics follows from the conservation of the energy--momentum tensor \cite{LL6}. Note also that both energy--momentum conservation and thermodynamical arguments apply to each one of the cosmic fluids individually \cite{CC}. In our case, we are interested in the ``fluid'' of vacuum fluctuations. For those, the pressure makes up the energy density acting in all three spatial dimensions such that $\epsilon_\mathrm{vac}=3p_\mathrm{vac}$ (Appendix A) or
\begin{equation}
p_\mathrm{vac} = \frac{\epsilon_\mathrm{vac}}{3} \,. 
\label{eq:pvac}
\end{equation}
However, with this pressure the energy density $\epsilon_\mathrm{vac}$ of Eq.~(\ref{eq:total}) does not satisfy the Friedmann equation (\ref{eq:f2}): it would violate the first law. We need to supplement the energy with another energy density $\epsilon_\Lambda$.  As it only takes care of the mismatch in energy--momentum conservation, it must not contribute to the right--hand side of Friedmann's equation (\ref{eq:f2}):
\begin{equation}
p_\Lambda = -\epsilon_\Lambda \,. 
\label{eq:plambda}
\end{equation}
We thus obtain for the total energy density
\begin{equation}
\partial_t(\epsilon_\mathrm{vac}+\epsilon_\Lambda) = - 4\epsilon_\mathrm{vac}H \,.
\label{eq:mismatch}
\end{equation}
Consider now the corresponding energy--momentum tensor [Eq.~(\ref{eq:Tiso})] $\mathrm{diag}(\epsilon_\Lambda,-p_\Lambda,-p_\Lambda,-p_\Lambda) = \epsilon_\Lambda \mathbb{1}$ according to Eq.~(\ref{eq:plambda}). This tensor has the trace $4\epsilon_\Lambda$, whereas for classical electromagnetic fields the trace of the energy--momentum tensor is zero \cite{LL2}. Renormalization causes a {\it trace anomaly} \cite{Wald}. 

Formula~(\ref{eq:mismatch})  is the defining equation of the anomalous energy density $\epsilon_\Lambda$. We see from our result (\ref{eq:total}) for $\epsilon_\mathrm{vac}$ that we can directly integrate it:
\begin{equation}
\epsilon_\mathrm{vac}+\epsilon_\Lambda= \epsilon_\infty + \frac{2\hbar}{3\pi^2c\ell^2} \, \dot{H} \,.
\label{eq:energyformula}
\end{equation} 
The integration constant $\epsilon_\infty$ is all that is left in the infinite future when the universe settles to exponential expansion \cite{MC} and $H=\mathrm{const}$. This integration constant is the cosmological constant \cite{Einstein}. 
 
 \section{Cosmic expansion}  
 
 \subsection{Gravity}
 
 The only force acting over the vast distances of space is gravity. The content of the universe --- radiation, matter and the quantum vacuum --- generates a gravitational background field that, in turn, decelerates the cosmic expansion --- if the force is attractive. This background field is weak and almost non--relativistic \cite{LL2}. We can use Newton's law to work it out. Let us briefly review the known theory \cite{Kolomeisky,MC} before we show how the quantum vacuum enters the picture. 
 
\begin{figure}[h]
\begin{center}
\includegraphics[width=20pc]{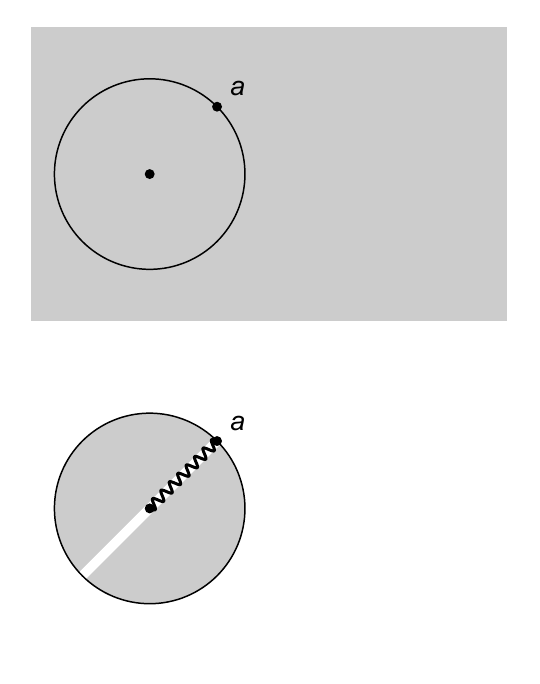}
\caption{
\small{Gauss' law. Top: Pick an arbitrary point (dot) in the homogeneous and isotropic universe (gray background). Take another point the distance $a$ away. The gravitational force of the background on the second point, relative to the first one, tells how the universe is accelerating. Bottom: According to Gauss' law the gravitational acceleration of the second point, with respect to the first one, depends only on the interior of the sphere around the first point. There gravity acts like the restoring force of an oscillator (spring). As the density falls with $a$ this is a nonlinear oscillator, and it may even reverse sign if the pressure is sufficiently negative, Eq.~(\ref{eq:einstein}). 
}
\label{fig:gauss}}
\end{center}
\end{figure}
 
Consider a sphere of radius $a$ in the homogeneous and isotropic universe (Fig.~\ref{fig:gauss}). According to Gauss' law the flux of the gravitational field across the surface of this sphere is given by $4\pi G$ times the integral of the density $\rho$ (where $G$ denotes Newton's gravitational constant). As the universe is homogeneous and isotropic over cosmological scales \cite{Survey} the flux is isotropic and the density homogeneous. The force of the gravitational field gives the acceleration $\ddot{a}$. We divide the flux by the volume, note that the surface of the unit sphere is $4\pi$ and the volume $\frac{4\pi}{3}$, and obtain for the gravitational acceleration \cite{LL2}:
\begin{equation}
\frac{\ddot{a}}{a} = - \frac{4\pi}{3} \,G\rho \,.
\label{eq:newton}
\end{equation} 
In this Newtonian picture, mass points the distance $a$ away from the center of the sphere get decelerated by the total mass of the sphere; in the Einsteinian picture, the distance itself gets decelerated: the measure of length is modified for everything, including the electromagnetic field. Apart from this conceptual difference, relativity adds two quantitative features \cite{LL2}: all the energy density $\epsilon$ gravitates, not only the rest mass, and also the pressure $p$. Since the pressure acts in all three spatial dimensions, $p$ receives a factor of three such that we get instead of the Newtonian Eq.~(\ref{eq:newton}) the Einsteinian
\begin{equation}
\frac{\ddot{a}}{a} = - \frac{4\pi G}{3c^2} \,(\epsilon+3p) \,.
\label{eq:einstein}
\end{equation} 
The gravitational force is attractive if $p>-\epsilon/3$, which is the case for matter and radiation, but according to Eq.~(\ref{eq:plambda}) the anomaly generates a repulsive force and may cause the cosmic expansion to accelerate. Astronomical observations \cite{Supernovae1,Supernovae2} have indeed shown that the expansion is accelerating. 

One can integrate \cite{CC} the dynamic equation (\ref{eq:einstein}) combined with the thermodynamic relation (\ref{eq:f2}). One obtains for the Hubble parameter (\ref{eq:hubble}):
\begin{equation}
H^2 + \frac{k}{a^2} = \frac{8\pi G}{3c^2}\,\epsilon \quad\mbox{with}\quad k=\mathrm{const}.
\label{eq:f1}
\end{equation} 
This is the other of Friedmann's equations \cite{Friedmann1,Friedmann2}. The integration constant $k$ quantifies the spatial curvature of the universe \cite{LL2}. CMB data \cite{Planck} have shown that 
\begin{equation}
k \sim 0 \,.
\label{eq:flat}
\end{equation} 
Space is flat to an excellent approximation. This concludes our miniature review of cosmic dynamics \cite{MC}. 

\subsection{Quantum buoyancy}

Consider now the influence of the vacuum energy on the cosmic dynamics. We have derived formula (\ref{eq:energyformula}) for the total of the vacuum--energy densities, $\epsilon_\mathrm{vac}+\epsilon_\Lambda$. Let us write this formula for the corresponding mass densities in light of Friedmann's equation (\ref{eq:f1}):
\begin{equation}
\rho_\mathrm{vac}+\rho_\Lambda=\frac{\epsilon_\mathrm{vac}+\epsilon_\Lambda†}{c^2}=\rho_\infty + 2\kappa \, \frac{3}{8\pi G}\, \dot{H} 
\label{eq:massformula}
\end{equation} 
with $\kappa$ given by Eqs.~(\ref{eq:planck}) and (\ref{eq:kappa}). In cosmology, time is typically measured in terms of the redshift \cite{MC} $z=a^{-1}-1$, {\it i.e.} in terms of the scale factor $a$. Let us describe the evolution as a function of $a$. From Eq.~(\ref{eq:hubble}) follows 
\begin{equation}
\dot{H} = H a \partial_a H = \frac{1}{2} a \partial_a H^2 \,.
\label{eq:dotH}
\end{equation} 
Let $\rho_\mathrm{M}$ denote the total density of matter (baryonic and dark) and $\rho_\mathrm{R}$ the density of radiation (photons and neutrinos). The total density is $\rho=\rho_\mathrm{R}+\rho_\mathrm{M}+\rho_\mathrm{vac}+\rho_\Lambda$. We obtain from Friedmann's equation (\ref{eq:f1}) with zero curvature $k$, and Eqs.~(\ref{eq:massformula}) and (\ref{eq:dotH}):
\begin{equation}
\rho-\kappa\, a\partial_a\rho = \rho_\mathrm{R}+\rho_\mathrm{M}+\rho_\infty \,.
\label{eq:diff}
\end{equation} 
This is an inhomogeneous linear differential equation for $\rho$. The homogeneous solution would be proportional to $\sqrt[\kappa]{a}$ and hence grow indefinitely for $a\rightarrow\infty$. This is impossible for a density. We thus put the homogeneous solution to zero and consider only the inhomogeneous one. As the differential equation (\ref{eq:diff}) is linear, each fluid is influenced by the quantum vacuum individually. This is a highly nontrivial aspect of the renormalized vacuum energy specific to our result (\ref{eq:total}). No other hypothetical vacuum energy would have this feature.

Radiation and matter are characterized by a simple equation of state \cite{MC} that relates the pressure to the energy density: 
\begin{equation}
p_m=w_m\epsilon_m \quad\mbox{with}\quad w_m=\mathrm{const}.
\label{eq:eos}
\end{equation} 
For radiation $w_\mathrm{R}=1/3$ (Appendix A) while for matter $w_\mathrm{M}=0$ (and for the cosmological constant $w_\Lambda=-1$). From Friedmann's thermodynamical relation (\ref{eq:f2}) follows 
\begin{equation}
\epsilon_m \propto a^{-3(1+w_m)} \,.
\label{eq:scaling}
\end{equation} 
Let $\varrho_m$ denote the contribution of each individual density $\rho_m$ to the total density $\rho$, including the vacuum contribution generated. We solve Eq.~(\ref{eq:diff}) for $\rho=\varrho_m$ with $\rho_m=\epsilon_m/c^2$ of Eq.~(\ref{eq:scaling}) as source,
\begin{equation}
\varrho-\kappa\, a\partial_a\varrho = \rho_m \,,
\end{equation}
set the homogeneous solution to zero, and obtain
\begin{equation}
\varrho_m = \frac{\rho_m}{1+3(1+w_m)\kappa} \,.
\label{eq:buoyancy}
\end{equation} 
This is the effective gravitational density of the fluid. There the original density $\rho_m$ is reduced by the characteristic factor $1+3(1+w_m)\kappa$ as if the fluid were partially buoyant in the quantum vacuum. 

\subsection{Hubble tension} 

How would our result (\ref{eq:buoyancy}) appear in cosmological data? The most precise data in cosmology have been obtained from measurements of the fluctuations of the CMB \cite{Planck}. These fluctuations were modulated by sound waves in the early universe when the baryonic matter was sufficiently ionized to couple strongly to light \cite{MC}. Sound waves are made by two counteracting forces: pressure and inertia. Light provided the pressure and baryonic matter, together with light, the inertia \cite{CC}. The densities of light and matter producing those sound waves are not affected by the quantum vacuum, but the expansion should be, according to our theory. This has consequences on the scale of the correlation spectrum. 

\begin{figure}[h]
\begin{center}
\includegraphics[width=18pc]{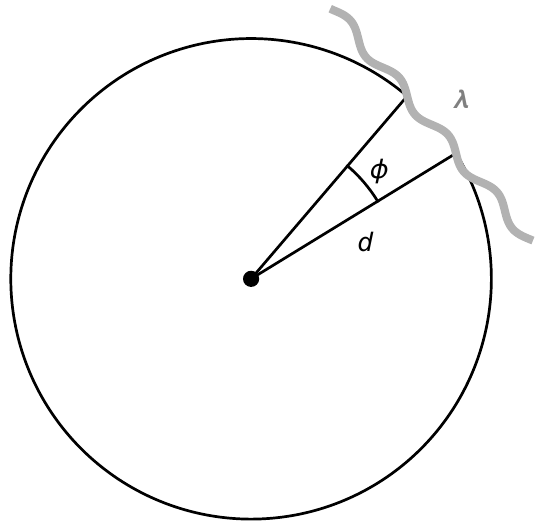}
\caption{
\small{Angular versus acoustical scale in the CMB. The CMB consists of thermal radiation with average temperature $T_0$ and variations $\delta T$. The variations were modulated by waves with wavelength $\lambda$ when the CMB was released. They are observed as angular variations of temperature. The original temperature undulations depend on all cosmic parameters, except the cosmological constant $\Lambda$, because $\Lambda$ did not play a significant role at the time of emission. But, in order to relate the angle $\phi$ to the wavelength $\lambda$ one needs to know the distance $d$. As light travels in conformal time $\tau$ like in free space, $d=c\tau$, which does depend on the cosmological constant, see Eqs.~(\ref{eq:distance}) and (\ref{eq:HlambdaCDM}) or (\ref{eq:hubblemodified}). This is how $\Lambda$ is inferred from CMB measurements \cite{Planck}.
}
\label{fig:angle}}
\end{center}
\end{figure}

The measured quantities are temperature variations as a function of spherical angle, but they originate from waves at the time the CMB was released (called the time of last scattering \cite{MC}) at which the universe was about $10^{-3}$ times smaller than it is today \cite{MC}. In order to relate angles $\phi$ to wavelengths $\lambda$ one needs to know the distance $d$ (Fig.~\ref{fig:angle}):
\begin{equation}
\phi\sim\frac{\lambda}{d} \,.
\label{eq:phi}
\end{equation} 
This distance is not simply given by $c$ times the time of last scattering ($13.8\mathrm{Gy}$) because the universe has expanded and hence increased the distance, or equivalently, the speed of light has varied as $c/n$ with the refractive index $n$ given by the scale factor, Eq.~(\ref{eq:a}). As light propagates with conformal time [Eq.~(\ref{eq:tau0})] like in empty Minkowski space, the distance traveled is $c$ times the conformal time, which amounts to about $46.2\mathrm{Gly}$ for the actual measured parameters of the  universe \cite{CC}. 

We denote the scale factor at the time of last scattering by $a_*$ and have according to Eqs.~(\ref{eq:tau0}), (\ref{eq:a}) and (\ref{eq:hubble}):
\begin{equation}
d = c\int_{a_*}^1\frac{da}{a^2H}
\label{eq:distance}
\end{equation} 
where $H$ is given by the Friedmann equation (\ref{eq:f1}). In the standard model of cosmology, the $\Lambda$ Cold Dark Matter  ($\Lambda$CDM) model, we write the densities $\rho_m$ in the form of the Friedmann equation (\ref{eq:f1}) with scaling law (\ref{eq:scaling}) as
\begin{equation}
\frac{8\pi G}{3}\,\rho_m = \frac{H_0^2\,\Omega_m}{a^{3(1+w_m)}} \quad\mbox{for}\quad m\in\{\mathrm{R},\mathrm{M},\Lambda\}
\label{eq:lambdaCMB}
\end{equation} 
with constant rate $H_0$, dimensionless constants $\Omega_m$ and the $w_m$ from the equation of state (\ref{eq:eos}): $w_\mathrm{R}=1/3$, $w_\mathrm{M}=0$ and $w_\Lambda=-1$. We thus have for the total Hubble parameter:
\begin{equation}
H^2=H_0^2\left(\frac{\Omega_\mathrm{R}}{a^4} +\frac{\Omega_\mathrm{M}}{a^3} + \Omega_\Lambda \right) .
\label{eq:HlambdaCDM}
\end{equation} 
We require that the $\Omega_m$ quantify the relative weights of the cosmic constituents radiation, matter and the cosmological constant, which implies
\begin{equation}
\Omega_\mathrm{R}+\Omega_\mathrm{M}+\Omega_\Lambda = 1 \,.
\end{equation} 
From this and Eq.~(\ref{eq:HlambdaCDM}) follows that the constant $H_0$ is the Hubble parameter at the present time when $a=1$, the Hubble constant ---- provided of course the $\Lambda$CDM model holds. 

In our case, we obtain from Eq.~(\ref{eq:buoyancy}) in the Friedmann equation (\ref{eq:f1}) the modified Hubble parameter
\begin{equation}
\frac{H^2}{H_0^2}=\frac{\Omega_\mathrm{R}}{(1+4\kappa)a^4} +\frac{\Omega_\mathrm{M}}{(1+3\kappa)a^3} + \Omega_\infty \, .
\label{eq:hubblemodified}
\end{equation} 
Note that this effective modification of the $\Lambda$CDM model only appears in the cosmic dynamics; the densities $\rho_m$ themselves are not affected by the quantum vacuum. Now, during the time the CMB was formed, the cosmological constant is negligible compared with the other densities \cite{CC} such that the dynamics is only governed by $H_0^2\Omega_\mathrm{R}$ and $H_0^2\Omega_\mathrm{M}$. Furthermore,  $H_0^2\Omega_\mathrm{R}$ is given by the Stefan--Boltzmann law \cite{CC}. We may thus assume, to a good approximation, that the shape of the CMB correlation curve is not influenced by the quantum vacuum and use the same parameters $H_0$, $\Omega_\mathrm{R}$ and $\Omega_\mathrm{M}$ as obtained from CMB data \cite{Planck}. 

Although the shape of the CMB curve is hardly affected, the scaling is, because the modified Hubble parameter [Eq.~(\ref{eq:hubblemodified})] alters the conformal distance [Eq.~(\ref{eq:distance})] that relates, via Eq.~(\ref{eq:phi}), the angular scale to the acoustical scale (Fig.~\ref{fig:angle}). Therefore, the value of $\Omega_\infty$ in Eq.~(\ref{eq:hubblemodified}) will differ from $\Omega_\Lambda$ in Eq.~(\ref{eq:HlambdaCDM}). The requirement
\begin{equation}
d = d_{\Lambda\mathrm{CDM}}
\label{eq:dcondition}
\end{equation} 
relates the two. For calculating $d_{\Lambda\mathrm{CDM}}$ and $d$ we may put $\Omega_\mathrm{R}=0$, because, during most of the cosmic expansion since the release of the CMB, the contribution of radiation to gravity has been negligible \cite{CC}.  As $a_*\approx 10^{-3}$ we may put it to zero as well. We obtain from Eq.~(\ref{eq:distance}):
\begin{equation}
d_{\Lambda\mathrm{CDM}} = \frac{2c}{H_0\sqrt{\Omega_\mathrm{M}}}\, {}_2 F_1\left(\frac{1}{6}, \frac{1}{2}, \frac{7}{6}, -\frac{\Omega_\Lambda}{\Omega_\mathrm {M}}\right)
\label{eq:dlambdaCDM}
\end{equation}
in terms of Gauss' hypergeometric function. For our case (\ref{eq:hubblemodified}) we only need to replace $\Omega_\mathrm{M}$ by $\Omega_\mathrm{M}/(1+3\kappa)$ and $\Omega_\Lambda$ by $\Omega_\infty$, and get
\begin{equation}
d = \frac{2c\sqrt{1+3\kappa}}{H_0\sqrt{\Omega_\mathrm{M}}}\, {}_2 F_1\left(\frac{1}{6}, \frac{1}{2}, \frac{7}{6}, -(1+3\kappa)\,\frac{\Omega_\infty}{\Omega_\mathrm {M}}\right) .
\end{equation}
We have argued (Sec.~I) that $\ell\gtrsim2\ell_\mathrm{P}$ is a good assumption for a cut--off, which implies [Eq.~(\ref{eq:kappa})] that $\kappa\approx 10^{-2}$. For each $\kappa$ we solve Eq.~(\ref{eq:dcondition}) numerically, using the values for the $\Omega_m$ from Planck satellite data \cite{Planck} ($\Omega_\mathrm{M}=0.3153$ and $\Omega_\Lambda=0.6847$). For $\Omega_\mathrm{R}=0$ we get from Eq.~(\ref{eq:hubblemodified}) the Hubble constant
\begin{equation}
\left. H \right|_{a=1} = \sqrt{\Omega_\infty + \frac{\Omega_\mathrm{M}}{1+3\kappa}} \,.
\end{equation}
We obtain the measured value \cite{Riess} $\left. H \right|_{a=1} = 1.084H_0$ for the parameter $\kappa=1.28\times 10^{-2}$ that corresponds [Eq.~(\ref{eq:kappa})] to the cut--off length $\ell=4.69\ell_\mathrm{P}$. In this case $\Omega_\infty=0.871$, which significantly deviates from $\Omega_\Lambda$. With the expected order of magnitude for the cut--off length we have thus reproduced the deviation of the Hubble constant. Other, more subtle tensions \cite{Abdalla} might also be resolved with our theory, which requires the numerical solution of the relativistic Boltzmann equations for the CMB \cite{MC} and is beyond the scope of this paper. Yet our simple, approximate result already indicates that one might actually resolve the most urgent problem of astrophysics, the cosmological tensions \cite{Abdalla}, with ideas from condensed--matter physics.

 \section{Conclusion}
 
 We have considered spatially uniform, time--dependent media with equal electric and magnetic response, Eq.~(\ref{eq:epsmu}). In such media, classical electromagnetic waves propagate like in empty, flat space with time transformed \cite{GREE} to the conformal time $\tau$ defined in Eq.~(\ref{eq:tau0}). Quantum fluctuations propagate with conformal time, too, and are correlated like in empty space with transformed time $\tau$, Eqs.~(\ref{eq:Kresult}) and (\ref{eq:s2}). In empty, flat space the (renormalized) vacuum energy is exactly zero. Yet, as we have argued, in time--dependent media, the energy densities depend on the time evolution, because the renormalizing Green function must not remember nor anticipate the full evolution, the full conformal time $\tau$, but may only depend on the present electromagnetic response and its first and second derivative \cite{Itay,Itai,Horner}. Renormalization is local \cite{London} --- in our case, local in time. 

Causality implies that the dielectric environment of the renormalizing Green function depends on the earlier of its two times. Causality combined with locality, Eqs.~(\ref{eq:H1}) and (\ref{eq:H2}), creates a discontinuity in the third derivative of the parameter (\ref{eq:eta}) that prevents the application of the standard fluctuation--dissipation theorem (Sec.~IIB). We can no longer directly infer \cite{Lifshitz} the correlations from the fluctuations, but need to transform time on the complex plane (Sec.~IIIC). The resulting energy density tracks the time evolution, Eq.~(\ref{eq:total}), and becomes significant for nearly dispersionless media. 

Such a dispersionless medium with equal electric and magnetic response is the ``medium'' of space in cosmology \cite{MC}, because flat space expanding in time with scale factor $a(t)$ is equivalent \cite{Plebanski} to the medium of Eq.~(\ref{eq:epsmu}) with refractive index $n=a$ (Appendix A). According to the equivalence principle, this medium is dispersionless until the Planck scale of Eq.~(\ref{eq:planck}).

Thermodynamics or, equivalently \cite{LL6}, energy--momentum conservation adds another twist to the vacuum energy: a trace anomaly \cite{Wald}. The renormalized energy--momentum tensor of the electromagnetic fluctuations is no longer traceless ($\epsilon_\mathrm{vac}-3p_\mathrm{vac}=0$),  but contains an additional component of energy density $\epsilon_\Lambda$ and pressure $p_\Lambda$ with $p_\Lambda=-\epsilon_\Lambda$. This is the equation of state of the cosmological constant \cite{Einstein}, but note that $\epsilon_\Lambda$  is not necessarily constant. The total $\epsilon_\mathrm{vac}+\epsilon_\Lambda$ tracks the time derivative of the Hubble parameter, Eq.~(\ref{eq:energyformula}), while Einstein's cosmological constant \cite{Einstein} emerges as an integration constant.

Cosmic expansion, or equivalently, the time--dependance of the medium, Eq.~(\ref{eq:epsmu}), generates vacuum energies and pressures that then, due to their gravity, act back on the cosmic evolution. In the relativistic Gauss' law of gravity, Eq.~(\ref{eq:einstein}), the negative pressure $p_\Lambda$ causes a repulsive contribution that may overwhelm the gravitational attraction and accelerate the cosmic expansion, as has been observed \cite{Supernovae1,Supernovae2}. Although the vacuum responds as a whole to the various cosmic constituents, and it is their combined gravity that  influences the evolution, we found a remarkable simplification: each constituent behaves as if its weight is reduced by a characteristic buoyancy factor, Eq.~(\ref{eq:buoyancy}). 

The quantum buoyancy of the vacuum modifies the cosmic evolution, Eq.~(\ref{eq:hubblemodified}), such that the Hubble constant --- the Hubble parameter (\ref{eq:hubble}) at the present time --- deviates from its projection without buoyancy by as much as $10\%$, in agreement with astronomical measurements \cite{Riess}. Our theory does not definitely resolve the Hubble tension, as it relies on an unknown coupling parameter that depends on the dispersion length $\ell$, but for $\ell$ in the order of twice the Planck length $\ell_\mathrm{P}$ we match the measurements \cite{Riess}. The physics on the Planck scale is unknown. However, it is reasonable to assume \cite{Jacobsondispersion} that a potential discreteness of space at $2\ell_\mathrm{P}$ appears as dispersion for lengths exceeding $2\ell_\mathrm{P}$ (in analogy to the dispersion of phonons in a crystal lattice \cite{Essential}). The dispersion of light propagation near the Planck scale does also resolve the trans--Planckian problem \cite{Jacobsondispersion} of Hawking radiation \cite{Bekenstein,Hawking1,Hawking2,Brout} which has been the focus of many analogues of gravity \cite{Unruh81,Volovik,Visser,Unsch,Faccionotes,Kolomeisky,Jacquet} in experiments \cite{Weinfurtner,Euve,Steinhauer,Drori}.

We have exclusively focused on the electromagnetic field, as all the experimental evidence for vacuum forces comes from quantum electromagnetism \cite{Rodriguez,Benea,Settembrini}. Tests of our specific theory --- other then in astronomy --- could perhaps be done with quantum fluctuations of the electromagnetic field as well. Although condition (\ref{eq:epsmu}) of equal electric and magnetic response is difficult to achieve in the laboratory \cite{SSS}, one might probably observe similar effects from a time--dependent electric and constant magnetic response, in addition to the particle creation known \cite{Dodonov} and measured \cite{Wilson,Hakonen,Veccoli} as the dynamical Casimir effect.

Which other fields could contribute to the vacuum energy in space and how would they affect it? \cite{Zeldovich} Fermions polarize the vacuum due to virtual particle--antiparticles pairs, and their polarization energy might contribute to the cosmological constant \cite{Tkatchenko}. The fields of other elementary bosons \cite{Weinfield} --- except the Higgs boson ---- are gauge fields, constructed from the same recipe as the electromagnetic field, but with mass (for the weak interaction) or nonlinearity (important for the strong interaction). In case they also contribute to the cosmological vacuum energy, one may trivially include them in our coupling parameter $\kappa$, simply by multiplying $\kappa$ by the number of fields and enlarging the cut--off by the square root of that number. The $\kappa$ parameter contains some information on the physics near the Planck scale, but not all, and in practice, $\kappa$ should be fitted to astronomical data.

If our theory of quantum noise in time--dependent media, developed in condensed matter physics, is indeed applicable to cosmology, it would relate the physics near the smallest of scales with observations on the largest scale --- the Planck scale written on the stars. 

\section*{Acknowledgements}

We are grateful for discussions with
Ofer Aharony,
Viktar Asadchy,
Dror Berechya,
Michael Berry,
Kfir Blum,
Nikolay Ebel,
Eren Erkul,
Mathias Fink, 
Uwe Fischer, 
Helmut H\"{o}rner,
Jonathan Kogman,
Amaury Micheli, 
Lukas Rachbauer,
Scott Robertson,
Stefan Rotter,
William Simpson,
Alexandre Tkatchenko,
Grisha Volovik,
Robert Wald,
Eli Waxman,
Chris Westbrook,
and Anton Zeilinger.
Our paper was supported by the Murray B. Koffler Professorial Chair of the Weizmann Institute of Science.


\appendix

\section{Brief excursion into general relativity}

While the main body of this paper does not rely on general relativity, but applies to spatially uniform dielectric media with equal electric and magnetic response, and negligible dispersion, it is instructive to see where the ``medium of space'' comes from \cite{GREE}. Modern cosmology \cite{MC,CC} starts from the Friedmann--Lema\^{i}tre--Robertson--Walker metric with zero spatial curvature:
\begin{equation}
ds^2 = c^2dt^2 - a^2d{\bm r}^2 \,.
\label{eq:metric}
\end{equation}
The metric (\ref{eq:metric}) describes the infinitesimal increment $ds/c$ of proper time \cite{LL2} perceived along any timelike trajectory. Space is assumed to be flat such that it enters with Euclidean line element $d{\bm r}^2=dx^2+dy^2+dz^2$ in Cartesian coordinates, but the measure of length is multiplied by the time--dependent scale factor $a$. The metric (\ref{eq:metric}) is a quadratic form $g_{\alpha\beta}\,dx^\alpha dx^\beta$ of the coordinate increments $dx^\alpha=(cdt,dx,dy,dz)$. Throughout this Appendix we adopt Einstein's summation convention over repeated upper and lower indices. The matrix $g_{\alpha\beta}$ is called the metric tensor with, in our case:
\begin{equation}
g_{\alpha\beta}= \mathrm{diag}(1,-a^2,-a^2,-a^2) \,.
\end{equation}
We get for the matrix inverse $g^{\alpha\beta}$ of the metric tensor
\begin{equation}
g^{\alpha\beta}= \mathrm{diag}(1,-a^{-2},-a^{-2},-a^{-2})
\label{eq:ginv}
\end{equation}
and for its determinant 
\begin{equation}
g=\det(g_{\alpha\beta}) = -a^6 \,.
\label{eq:gdet}
\end{equation}
The determinant of the metric accounts for the space--time volume element as $\sqrt{-g}\, d^4x$ (with $\sqrt{-g}=a^3$ in our case). 

The classical electromagnetic field is described by the four--potential $A_\alpha$ that comprises the electric potential $V$ and the magnetic vector potential ${\bm A}$ as $A_\alpha=(V,-c{\bm A})$. The potentials generate the electric field strength $\bm{E}=-\nabla V - \partial_t \bm{A}$ and the magnetic induction $\bm{B}=\nabla\times\bm{A}$ that constitute the field--strength tensor 
\begin{equation}
F_{\alpha\beta} = \partial_\alpha A_\beta - \partial_\beta A_\alpha
\label{eq:ftensor}
\end{equation}
which reads in matrix form:
\begin{equation}
F_{\alpha\beta} = 
\begin{pmatrix}
0 & E_x & E_y & E_z \\
-E_x &  0 & -cB_z & cB_y \\
-E_y &  cB_z & 0 & -cB_x \\
-E_z &  -cB_y & cB_x &  0
\end{pmatrix}
.
\label{eq:fieldtensor}
\end{equation}
If we replace the four--potential $A_\alpha$ by $A_\alpha+\partial_\alpha\varphi$  in Eq.~(\ref{eq:ftensor}) we get the same $F_{\alpha\beta}$, which means the field--strength tensor is gauge--invariant. Being constructed from potentials, the fields naturally satisfy the first group of Maxwell's equations \cite{Jackson}:
\begin{equation} 
\nabla\cdot\bm{B} = 0 \,,\quad \nabla\times\bm{E} + \partial_t\bm{B} = \bm{0} \,.
\end{equation}
The second group of Maxwell's equations turns out to follow from minimising the action \cite{LL2}
\begin{equation} 
S = \frac{1}{c} \int \mathscr{L} \sqrt{-g}\,d^4 x
\label{eq:action}
\end{equation}
with the electromagnetic Lagrangian 
\begin{equation} 
\mathscr{L}_\mathrm{EM} = -\frac{\varepsilon_0}{4} \,F_{\alpha\beta} F^{\alpha\beta} 
\label{eq:EMlagrangian}
\end{equation}
where $F^{\alpha\beta}$ is defined as
\begin{equation} 
F^{\alpha\beta}=g^{\alpha\mu}g^{\beta\nu} F_{\mu\nu} \,.
\label{eq:fdef}
\end{equation}
The Lagrangian $\mathscr{L}_\mathrm{EM}$ is the only scalar quadratic in the field strengths \cite{LL2} (up to the constant prefactor $\varepsilon_0$). Therefore, the Euler--Lagrange equations are the only linear field equations derivable from a four--potential that are gauge--invariant and consistent with general relativity. 

Note also that the action (\ref{eq:action}) with Lagrangian (\ref{eq:EMlagrangian}) is conformally invariant: Suppose we replace the metric tensor $g_{\alpha\beta}$ by $\Omega^2g_{\alpha\beta}$ with an arbitrary function $\Omega$ of the coordinates. In this case $F^{\alpha\beta} \sqrt{-g}$ is unchanged and so is the action. In our cosmological model, we may express the metric (\ref{eq:metric}) in terms of the conformal time [Eq.~(\ref{eq:tau0})] and get 
\begin{equation}
ds^2 = a^2(c^2d\tau^2 - d{\bm r}^2)
\end{equation}
such that $\Omega=a$. From the conformal invariance of the electromagnetic action follows that classical electromagnetic waves in the space--time geometry (\ref{eq:metric}) or, equivalently, in spatially uniform materials with equal electric and magnetic response, propagate as in free space with transformed time. 

The electromagnetic field equations are the Euler--Lagrange equations \cite{LL2} derived from the action (\ref{eq:action}) with the Lagrangian (\ref{eq:EMlagrangian}):
\begin{equation} 
0 = \frac{\partial\sqrt{-g}\,\mathscr{L}_\mathrm{EM}}{\partial A_\beta} - \partial_\alpha \frac{\partial\sqrt{-g}\,\mathscr{L}_\mathrm{EM}}{\partial(\partial_\alpha A_\beta)} = \partial_\alpha \mathscr{H}^{\alpha\beta} 
\label{eq:maxwell}
\end{equation}
where we have defined the tensor density 
\begin{equation} 
\mathscr{H}^{\alpha\beta} = \varepsilon_0\sqrt{-g}\,F^{\alpha\beta} \,.
\label{eq:hdef}
\end{equation}
Let us represent $\mathscr{H}^{\alpha\beta}$ as 
\begin{equation} 
\mathscr{H}^{\alpha\beta} =
 \begin{pmatrix}
0 & -D_x & -D_y & -D_z \\
D_x &  0 & -H_z/c & H_y/c \\
D_y &  H_z/c & 0 & -H_x/c \\
D_z &  -H_y/c & H_x/c &  0
\end{pmatrix}
.
\label{eq:htensor}
\end{equation}
One verifies that Eq.~(\ref{eq:maxwell}) appears as the Maxwell equations in macroscopic media \cite{Jackson}:
\begin{equation} 
\nabla\cdot\bm{D} = 0 \,,\quad \nabla\times\bm{H} - \partial_t\bm{D} = \bm{0} \,.
\end{equation}
For a general space--time geometry, the constitutive equations relating $\bm{D}$ and $\bm{B}$ to $\bm{E}$ and $\bm{H}$ have been derived by P{\l}ebanski \cite{Plebanski}. In our case, Eq.~(\ref{eq:metric}), they are elementary to obtain: From definitions (\ref{eq:fdef}) and (\ref{eq:hdef}) with notations (\ref{eq:fieldtensor}) and (\ref{eq:htensor}) and expressions (\ref{eq:ginv}) and (\ref{eq:gdet}) we get Eq.~(\ref{eq:constitutive}). 

This result shows that the expanding universe with metric (\ref{eq:metric}) appears as a spatially uniform medium with equal electric and magnetic response, Eq.~(\ref{eq:epsmu}). The ``medium'' of empty space does neither depend on the energy nor on the wavenumber or frequency of electromagnetic radiation; the medium is linear and dispersionless, until classical general relativity loses its validity. This is a manifestation of the equivalence principle: all bodies fall the same way, including the ``body'' of the electromagnetic field, regardless of energy, wavenumber or frequency. In this paper, we assume the equivalence principle to hold until near the Planck scale of Eq.~(\ref{eq:planck}). Experience from vacuum forces in media \cite{Itay,Itai} and from the trans--Planckian problem of Hawking radiation in space \cite{Jacobsondispersion} suggests that ultimately dispersion will become relevant. In this paper, we describe dispersion in the most primitive way --- by a cutoff. As the physics near the Planck scale is unknown, it is probably wise to use such a simple model, with the cutoff fitted to astronomical data.

These data indicate how the universe has expanded. General relativity assumes that this expansion --- and all other gravitational phenomena --- are governed by the Lagrangian of the gravitational field \cite{LL2}:
\begin{equation} 
\mathscr{L}_\mathrm{G} = -\frac{c^4}{16\pi G}\,R 
\end{equation}
where $R$ denotes the space--time curvature scalar \cite{LL2} (the Ricci scalar). It is the only Lagrangian that depends on the space--time curvature and produces second--order field equations, consistent with Newtonian gravity in the non--relativistic limit \cite{LL2}. The prefactor, containing Newton's gravitational constant $G$, is chosen to fit this limit \cite{LL2}. The total Lagrangian $\mathscr{L}$ is the sum of $\mathscr{L}_\mathrm{G}$ and the Lagrangian $\mathscr{L}_\mathrm{M}$ of everything else, including the electromagnetic field. Minimizing the total action (\ref{eq:action}) with respect to the metric tensor turns out \cite{LL2} to give Einstein's field equations:
\begin{equation} 
R_\alpha^\beta - \frac{R}{2}\,\delta_\alpha^\beta = \frac{8\pi G}{c^4} T_\alpha^\beta 
\label{eq:einsteineqs}
\end{equation}
where $R_\alpha^\beta$ denotes the Ricci tensor \cite{LL2} and $T_\alpha^\beta$ the energy--momentum tensor \cite{LL2}:
\begin{equation} 
T_\alpha^\beta = 2g^{\beta\nu}\frac{\partial \mathscr{L}_\mathrm{M}}{\partial g^{\alpha\nu}} - \mathscr{L}_\mathrm{M}\,\delta_\alpha^\beta \,.
\label{eq:T}
\end{equation}
Independent of the laws of gravity, formula (\ref{eq:T}) also comes from the connection between symmetries and conservation laws. In classical mechanics\cite{LL1}, energy and momentum are the conserved quantities associated with the invariance of the equations of motion  to shifts in time and space. In general relativity\cite{LL2}, the equations of any field are generated by a Lagrangian $\mathscr{L}_\mathrm{M}$ that is invariant under general coordinate transformations, including shifts in $t$ and $\bm{r}$. From this more general invariance also follows \cite{LL2} Eq.~(\ref{eq:T}) --- apart from a constant prefactor. This factor is set by gravity, by the bare weight of energy and momentum. (In this paper, we have shown how the quantum vacuum alters the apparent weight of radiation and matter during cosmic expansion.) 

For the electromagnetic field with Lagrangian  (\ref{eq:EMlagrangian}) and (\ref{eq:fdef}) we immediately obtain from formula (\ref{eq:T}) the energy--momentum tensor:
\begin{equation} 
T_\alpha^\beta = \varepsilon_0 F^{\beta\nu}F_{\nu\alpha} + \frac{\varepsilon_0}{4} F_{\mu\nu}F^{\mu\nu} \delta_\alpha^\beta \,.
\label{eq:EMT} 
\end{equation}
In terms of the fields $\bm{E}$, $\bm{B}$ and $\bm{D}$, $\bm{H}$ we get from definitions (\ref{eq:fieldtensor}) and (\ref{eq:htensor}) the tensor density
\begin{equation} 
\sqrt{-g}\,T_\alpha^\beta = 
\begin{pmatrix}
\frac{1}{2}\bm{E}\cdot\bm{D} + \frac{1}{2}\bm{B}\cdot\bm{H}& -c \bm{D}\times\bm{B}\\[4pt]
\frac{1}{c} \bm{E}\times\bm{H} & \sigma
\end{pmatrix}
\label{eq:energymomentum}
\end{equation}
where $\sigma$ denotes the electromagnetic stress \cite{Jackson}
\begin{equation} 
\sigma = \bm{D}\otimes\bm{E} + \bm{B}\otimes\bm{H} - \left(\frac{\bm{D}\cdot\bm{E}}{2}+\frac{\bm{B}\cdot\bm{H}}{2}\right) \mathbb{1} \,.
\label{eq:stress}
\end{equation}
Expression (\ref{eq:energymomentum}) proves Eq.~(\ref{eq:uem}) for the energy densities of the electromagnetic field. For incoherent electromagnetic radiation,  the field strengths are not deterministic, but fluctuate statistically. For the quantum vacuum, these are quantum fluctuations. The averaged Poynting vector \cite{Jackson} $\langle\bm{E}\times\bm{H}\rangle$ and  momentum density \cite{Jackson} $\langle\bm{D}\times\bm{B}\rangle$ must vanish in an isotropic and homogeneous universe, for otherwise they would distinguish directions. The averaged stress must be isotropic, too, which implies that $\langle\sigma\rangle$ is proportional to the identity matrix times the pressure. We thus obtain for the averaged energy--momentum tensor itself:
\begin{equation} 
\langle T_\alpha^\beta\rangle = \mathrm{diag}(\epsilon, -p,-p,-p) 
\label{eq:Tiso}
\end{equation}
where $\epsilon$ denotes the proper energy density and $p$ the pressure. The proper energy density is the energy density divided by the measure $\sqrt{-g}$ of space--time volume,  in our case $a^3$, which gives Eq.~(\ref{eq:energies}). Pressure is the outward momentum flux from an infinitesimal volume element, and so it carries a minus sign in Eq.~(\ref{eq:Tiso}). There we have divided $-\langle\sigma\rangle$ by $a^3$ as well, because pressure has the physical units of an energy density and so should be treated like the proper energy density. For the electromagnetic field, the energy--momentum tensor is traceless, $T_\alpha^\alpha=0$, as one sees from Eq.~(\ref{eq:EMT}) and also from the explicit expressions (\ref{eq:energymomentum}) and (\ref{eq:stress}). For a traceless tensor of type (\ref{eq:Tiso}) we have $\epsilon=3p$, which gives Eq.~(\ref{eq:pvac}) for vacuum fluctuations. 

Note that the other ingredients of the cosmic mix \cite{MC,CC} --- radiation (photons and neutrinos), matter (baryonic and dark) and the cosmological constant --- must have energy--momentum tensors of the type (\ref{eq:Tiso}) as well, for being consistent with an isotropic and homogeneous universe. Their equation of state, Eq.~(\ref{eq:eos}), relates the pressure to the proper energy density. The cosmological constant we may put on the right--hand side of Einstein's equations (\ref{eq:einsteineqs}) and interpret it as a constant energy--momentum tensor. The only option for a constant $T_\alpha^\beta$ is to be proportional to the four--dimensional unity matrix $\delta_\alpha^\beta$, which implies Eq.~(\ref{eq:plambda}). 

Finally, we note that for an isotropic and homogeneous universe, the Einstein equations (\ref{eq:einsteineqs}) reduce \cite{LL2} to the Friedmann equations (\ref{eq:f2}) and (\ref{eq:f1}) where for the spatially flat case of Eq.~(\ref{eq:metric}) we have  $\mathrm{k}=0$. From the Friedmann equations (\ref{eq:f2}) and (\ref{eq:f1}) follows the Einsteinian modification (\ref{eq:einstein}) of the Newton equation (\ref{eq:newton}) of the universe. This was our starting point for calculating the effect of quantum noise in time--dependent media on the cosmic expansion. 




\begin{thebibliography}{99}
\bibitem{Atm}
The Casimir pressure between two perfect reflectors scales with the inverse forth power of the distance \cite{Rodriguez}, which needs to be in the order of $10\mathrm{nm}$ to reach $1\mathrm{atm}$.
\bibitem{DoS}
The Fourier--transformed Green function $\widetilde{G}_+$ is analytic in $\omega$ for $\mathrm{Im}\omega>0$. One finds from the Green equation that its poles are $\mp(\varepsilon_0c^2/\hbar)A_k({\bm r})A_k^*({\bm r}')/(\omega\mp\omega_k)$ in terms of the normalized \cite{Essential} stationary waves $A_k({\bm r})$ with eigenfrequencies $\omega_k$. From the analyticity of $\widetilde{G}_+$ follows $\mathrm{Im}\widetilde{G}_+({\bm r}, {\bm r}) =-\pi\,\mathrm{LDOS}$ for real $\omega>0$ with  $\mathrm{LDOS}=(\varepsilon_0c^2/\hbar) \sum_k|A_k({\bm r})|^2\delta(\omega-\omega_k)$.
\end{thebibliography}
\end{document}